# Towards a general characterization of flat fan sprays through Direct Numerical Simulations

Longxiang Huang[a,1], Benjamin Duret[b], François-Xavier Demoulin[b]

[a]*Institute for Aero Engine, Tsinghua University , China*
[b]*CNRS CORIA UMR 6614, University of Rouen Normandie, France*

**Abstract**

A numerical investigation of flat fan sprays is conducted via Direct Numerical Simulations (DNS). Diverging liquid sheets are generated using tailored initial velocity profiles, where the opening angle serves as an explicit control parameter. The analysis reveals two distinct regimes: at low Weber numbers, the sheet features thick, retracting rims moving at the Taylor-Culick velocity, though rim-driven break-up is not observed without advanced techniques like Adaptive Mesh Refinement (AMR). At high Weber numbers, aerodynamic instabilities govern disintegration, with hole break-up absent in all cases. Representing the spray as a triangular sheet, a simplified model is proposed to predict the axial thickness evolution, showing good agreement with numerical measurements. The study also quantifies the influence of Weber number and opening angle on surface wave properties. An existing break-up length model is successfully applied, incorporating the present initial conditions, offering a predictive tool for future numerical and experimental studies.

## 1. Introduction

Atomization and the development of sprays are important phenomena in a wide range of industrial applications such as the fuel injection in the combustion engine and agricultural spraying. The dynamics of liquid disintegration have been widely investigated using numerical approaches, with Direct Numerical Simulations (DNS) providing a detailed view of interfacial instabilities and break-up processes. For instance, Ménard et al. (2007) proposed a CLSVOF method to improve mass conservation in DNS of a round liquid jet's primary break-up. Jiao et al. (2017) conducted DNS for diesel injection with a realistic turbulence inlet profile. Asuri Mukundan et al. (2022b) first introduced a hybrid moment of fluidlevel set method to reconstruct the liquid/gas interface and adopted it for the detailed numerical simulation of an air-blasted liquid sheet (Asuri Mukundan et al., 2022a). The flapping mechanism of a liquid sheet induced by a high-speed stream has been examined (Odier et al., 2015), and significant research has also been dedicated to the DNS of crossflow configurations (Li and Soteriou, 2018; Chai et al., 2023; Mukundan et al., 2021; Behzad et al., 2016). Despite this broad coverage of different geometries, a specific configuration remains unexplored: to the authors' knowledge, no existing works in the literature tackle the DNS of a flat fan spray.

---

[1] Corresponding author. Email: huanglx@mail.tsinghua.edu.cn

A flat fan spray is generated when a liquid is pressurized through a fanshaped atomizer, producing a sheet that exhibits a characteristic divergence. After leaving the nozzle, the liquid sheet expands in the transverse direction while its thickness attenuates continuously in the axial direction as a consequence of mass conservation. This specific geometry is often generalized in the literature as an attenuating liquid sheet. The sheet eventually becomes susceptible to hydrodynamic instabilities, leading to its disintegration into ligaments and droplets. Recently, Sanadi (2022) has studied the thickness evolution of the flat fan spray through experimental approaches. Different break-up phenomenon have also been identified at various Weber number and Reynolds number. However, their work is limited to an experimental study and certain information such as the velocity distributions and the transverse thickness distributions cannot be easily obtained. The objective of the present work is to address this identified gap by conducting a numerical study of this particular spray type using Direct Numerical Simulations.

## 2. Disintegration of an attenuating liquid sheet

Research on flat fan sprays has historically unfolded in two main periods. The foundational work was established between 1950 and 1975, primarily by Dombrowski, Fraser, and their collaborators. In their pioneering studies (Dombrowski and Fraser, 1954; Dombrowski et al., 1960; Fraser et al., 1962), they applied linear stability analysis to systematically describe the formation, destabilization, and break-up mechanisms of these sheets, modeling the transition from a continuous sheet into ligaments and subsequently into droplets.

A more recent and active period of research extends from approximately 2000 to the present. During this time, Altimira et al. (2009) utilized numerical simulations to characterize fan spray atomizers, validating their results against experimental data (Rivas et al., 2005) and existing mathematical models. Complementing these numerical efforts, Altieri et al. (2014) provided further experimental details on the break-up processes and contributed to the mathematical modeling of liquid sheets. Building directly upon the early theoretical groundwork, Post and Hewitt (2018) developed a simplified model based on a previous analysis (Dombrowski et al., 1960) for predicting droplet sizes.

Across these studies, three primary disintegration modes for the flat fan spray have been consistently identified: the wave, hole, and rim break-up mechanisms (Fraser et al., 1962).

The wave break-up mechanism is primarily initiated by two types of disturbances: aerodynamic forces acting on the sheet surface, and irregularities at the nozzle exit that introduce turbulent perturbations. Dombrowski and Fraser (1954) systematically documented different sheet patterns under varying conditions and were among the first to establish this fundamental distinction.

Regarding aerodynamic instability, the foundational work is often attributed to Lord Rayleigh (Rayleigh, 1879), who identified two linearly independent instability modes. The first is the sinuous mode, characterized by the antisymmetric oscillation of the two interfaces while the sheet thickness remains approximately constant. The second is the varicose, or dilational mode, where the interfaces move symmetrically, leading to localized variations in sheet thickness.

Building upon this, Fraser et al. (1962) developed a break-up model based on the growth of sinuous waves. This model applies to the idealized scenario where a liquid sheet is subjected purely to aerodynamic forces, without significant disturbances from the nozzle that would otherwise lead to hole or bag break-up. In this process, an initial disturbance generates sinuous waves that amplify over time, producing a characteristic flapping motion. Once the wave amplitude reaches a critical threshold, ligaments are periodically stripped from the sheet in the transverse direction. These ligaments subsequently retract and break up into droplets via the Rayleigh-Plateau instability.

In addition to the aerodynamic waves that dominate the break-up of an attenuating liquid sheet, instabilities and disturbances originating from the nozzle also contribute to surface corrugations. When combined with the sheet's inherent transverse divergence and the external aerodynamic forces, these disturbances promote the formation of three-dimensional structures and enhance waviness in the direction normal to the sheet plane.

Fraser et al. (1962) observed this type of break-up in a single-phase liquid sheet under vacuum conditions, attributing it primarily to the impingement of droplets. Their explanation, however, did not fully account for other potential factors such as the release of dissolved gas, cavitation, or nozzle vibration. More recently, Asgarian et al. (2020) demonstrated experimentally that the ripples induced by droplet impingement are of smaller amplitude than those generated by nozzle turbulence and aerodynamic waves. In their study, internal nozzle turbulence and the resulting initial sheet disturbances were numerically triggered by incorporating a castellated mesh in the nozzle model.

In contrast to the aerodynamic waves that dominate the break-up of a liquid sheet, or the perforations that lead to interfacial irregularities, publications focusing directly on the rim of flat fan sprays have been limited since the period of Dombrowski and Fraser.

Dombrowski and co-workers investigated the behaviour of flat fan sprays in their series of works (Dombrowski and Fraser, 1954; Dombrowski et al., 1960; Fraser et al., 1962; Dombrowski and Johns, 1963; Crapper et al., 1973; Clark and Dombrowski, 1974; Dombrowski and Foumeny, 1998). They observed that upon leaving the nozzle, the sheet is bounded by two thick rims. These rims contract under the action of surface tension, forming a bell-shaped curve as the sheet develops downstream.

The break-up process of the rim initiates with the cylindrical rim subjected to the Rayleigh-Plateau instability, leading to the formation of varicose waves. These waves subsequently cause the development of liquid bulbs along the sheet's boundaries (Sanadi, 2022). These bulbous regions, possessing a significantly larger radius than the adjacent liquid sheet, are then ejected outwards under the influence of a centripetal force related to the rim's curved trajectory. The ejected liquid threads ultimately pinch off due to the Rayleigh-Plateau instability, forming a fish-bone structure whose morphology is largely determined by the Ohnesorge number of the liquid.

Fullana and Zaleski (1999) analysed the stability of a thin fluid sheet of uniform thickness bounded by a retracting rim using DNS. A span-wise perturbation with a large initial amplitude was introduced to trigger the instability. A large number of simulations with different physical parameters were performed, yet no break-up was observed in any scenario, as the growing rim radius consistently suppressed the instability. They concluded that perturbations are amplified only for very large wavelengths and over long periods, proposing a typical break-up criterion of 100 < $\lambda$ < 500, where $\lambda$ is the wavelength normalized by the rim radius.

Agbaglah et al. (2013) implemented Adaptive Mesh Refinement (AMR) and varied initial conditions to numerically achieve a configuration where fingers and droplets detach. They further pointed out that capturing this phenomenon requires a minimum of three mesh points within the liquid sheet and, given an aspect ratio of 0.2 between the sheet thickness and the rim radius, a minimum aspect ratio of 1500 between the wavelength and the grid size. This criterion renders rim-driven break-up extremely computationally expensive to simulate using standard DNS.

## 3. Numerical approach

### 3.1. DNS solver

The in-house code ARCHER is used throughout the thesis for various configurations. It is a highly structured and parallel code dedicated to solving the incompressible Navier-Stokes equations on a staggered Cartesian mesh:

$$\nabla \cdot \vec{u} = 0 \tag{1}$$

$$\frac{\partial \rho \vec{u}}{\partial t} + \rho(\vec{u} \cdot \vec{\nabla})\vec{u} = -\vec{\nabla}P + \vec{\nabla} \cdot \mu(\vec{\nabla}\vec{u} + \vec{\nabla}\vec{u}^T) + \vec{f} + \sigma\kappa\delta\vec{n} \tag{2}$$

where $u$ and $P$ are the velocity and pressure fields. $\rho$ is the density, $f$ is a source term, $\sigma$ is the surface tension, $\kappa$ is the interface curvature, $\delta$ is the Dirac function indicating the location of the interface, $n$ is the unit normal vector to the interface.

The numerical solver employs a projection method for velocity-pressure coupling (Chorin, 1997). The convective term is handled with an enhanced

Rudman's technique (Rudman, 1998; Vaudor et al., 2017), while the viscous term is computed following Sussman et al. (2007). Interface tracking is achieved using a coupled Level Set/Volume-of-Fluid (CLSVOF) method to ensure mass conservation (Ménard et al., 2007), with a Ghost Fluid Method applied to accurately capture interfacial pressure jumps (Fedkiw et al., 1999; Tanguy et al., 2007). The computations are performed on a staggered Cartesian (MAC) grid, where velocities are defined on a mesh offset by half a cell from the primary variable locations (Harlow et al., 1965).

This solver, ARCHER, has been extensively validated for atomization, spray, and liquid-gas flow studies (Martinez et al., 2021, 2023; Roa et al., 2023; Chéron et al., 2022; Deberne et al., 2024a,b; Duret et al., 2018). Originally developed by Ménard et al. (2007), its capabilities have been expanded over the past two decades. Key developments include the introduction of a homogeneous isotropic turbulence (HIT) configuration (Luret, 2010), its extension to evaporating flows (Duret, 2013), and the implementation of Lagrangian particle libraries (Chéron, 2020). For further details, the reader is referred to the cited literature.

*3.2. Numerical setup*

The internal structure of the flat fan spray nozzle has been properly set up to reproduce a realistic spray in previous works (Asgarian et al., 2020; Kashani et al., 2018). In particular, Asgarian et al. (2020) used a castellated mesh strategy at the nozzle level to trigger the initial turbulent environment, leading to irregularities and eventually holes and perforations on the liquid sheet.

Using our in-house DNS solver ARCHER, such a pre-processing strategy is not yet mature. Nevertheless, we used different velocity profiles as initial conditions to mimic the characteristic feature of divergence of the spray.

As a starting point, a Cartesian mesh is generated where the Injection boundary condition is set on the plane corresponding to $z = 0$, the Outflow boundary condition is set on the rest of the planes. A schematic of the mesh grid is presented in Fig. 1. The important parameters of the current configuration are summarized in Table. 1.

In experimental studies, the liquid sheet is perturbed at the nozzle level, and the axial velocity differs significantly between the edge and the center of the liquid sheet. Nevertheless, the axial velocity profile does not play a critical role in determining the overall patterns of the flat fan spray. Therefore, we assume that the axial velocity component $W$ along the z-axis or $X_3$ axis is constant and does not significantly influence the characteristics of the attenuating sheet. To create the 'flat fan' feature of a liquid sheet, the

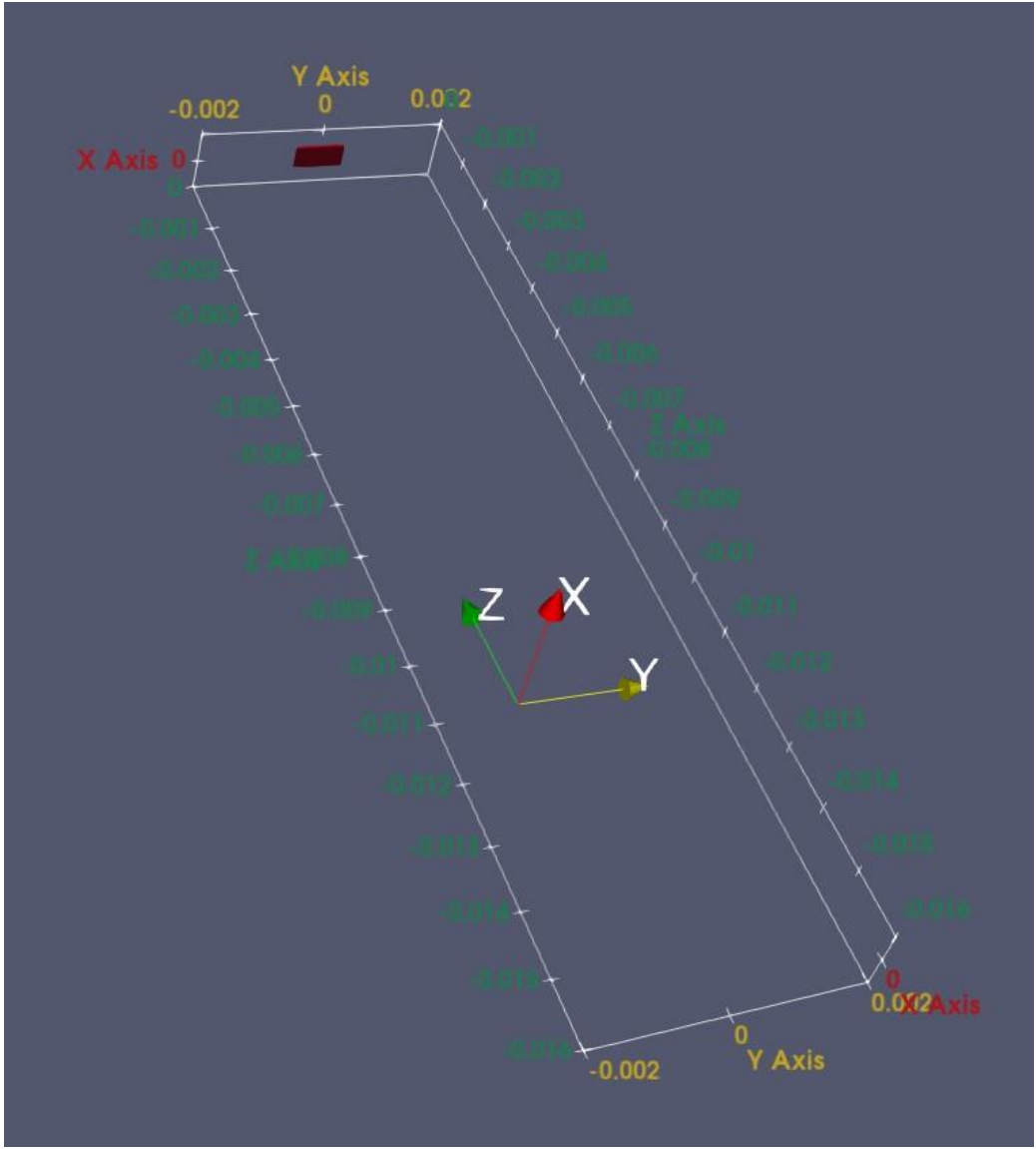

Figure 1: Schematic of the generated mesh grid. The initial zero level set is contoured in red.

Table 1: Parameters for current configuration

| Parameters | values (S.I. units) | |
|---|---|---|
| $\theta$ | 5° | |
| y | Transverse coordinates | |
| x | Normal coordinates | |
| Axial air velocity $W_g$ | 0 | |
| Axial liquid velocity $W_l$ | 80 | |
| Gas density $\rho_g$ | 25 | |
| Liquid density $\rho_l$ | 1000 | |
| Liquid dynamic viscosity $\mu_l$ | 0.001 | Liquid |
| Gas dynamic viscosity $\mu_g$ | $1.88 * 10^{-5}$ | Weber |
| Surface tension $\sigma$ | 0.07 | |
| Length of injector $L_{inj}$ | $4R_{inj} = 8 * 10^{-4}$ | |
| Width of injector $e_{inj}$ | $2R_{inj} = 4 * 10^{-4}$ | |
| Size indicator $R_{inj}$ | | |
| Lx (X3: normal direction) | $2 * 10^{-4}$ | |
| Ly (X2: transverse direction) | 0.0005(32 cells) 0.004(256 cells) | |
| Lz (X1: axial direction) | 0.016(1024 cells) | |
| dx (coarse mesh) | $1.5625 * 10^{-5}$ | |
| number ($We_l = \rho_l W_l^2 e_{inj}/\sigma$) | 36571 | |
| Liquid Reynolds number ($Re_l = \rho_l W_l e_{inj}/\mu_l$) | 32000 | |
| CFL | 0.2 | |
| Coarse mesh | 32*256*1024 | |
| Fine mesh | 64*512*2048 | |

following velocity profiles have been adopted:

$$W = \begin{cases} c & -R_{inj} \leq x \leq R_{inj} \quad \& \quad -2R_{inj} \leq y \leq 2R_{inj} \\ 0 & else \end{cases} \quad (3)$$

$$V = \begin{cases} \frac{WTan(\theta)y(j)}{2R_{inj}} & -R_{inj} \leq x \leq R_{inj} \\ \& \quad -2R_{inj} \leq y \leq 2R_{inj} \\ 0 & else \end{cases} \quad (4)$$

$$U = \begin{cases} \frac{-WTan(\theta)x(i)}{2R_{inj}} & -R_{inj} \leq x \leq R_{inj} \\ \& \quad -2R_{inj} \leq y \leq 2R_{inj} \\ 0 & else \end{cases} \quad (5)$$

In this formulation, non-zero values correspond to the liquid phase, while a value of zero is assigned to the gas phase. The variable $V$ represents the transverse velocity profile along the y-axis or $X_2$ axis, governing the expansion of the liquid sheet in the transverse plane. The variable $U$ denotes the normal velocity profile along the x-axis or $X_1$ axis, responsible for the attenuation of the liquid sheet thickness in the normal direction. The axial velocity component $W$ is set to a controllable constant value $c$. The opening angle $\theta$ is the parameter that controls the degree of divergence of the liquid sheet.

Even though the divergence-free condition is naturally handled and guaranteed once the simulation is started inside the solver for incompressible flow, the sum of derivatives of these velocity profiles concerning their respective axes equals zero to avoid any artificial errors at this stage.

For illustration purposes, using one set of the parameters, the transverse and normal velocity profiles are shown on the contour of the zero level set in Fig. 2.

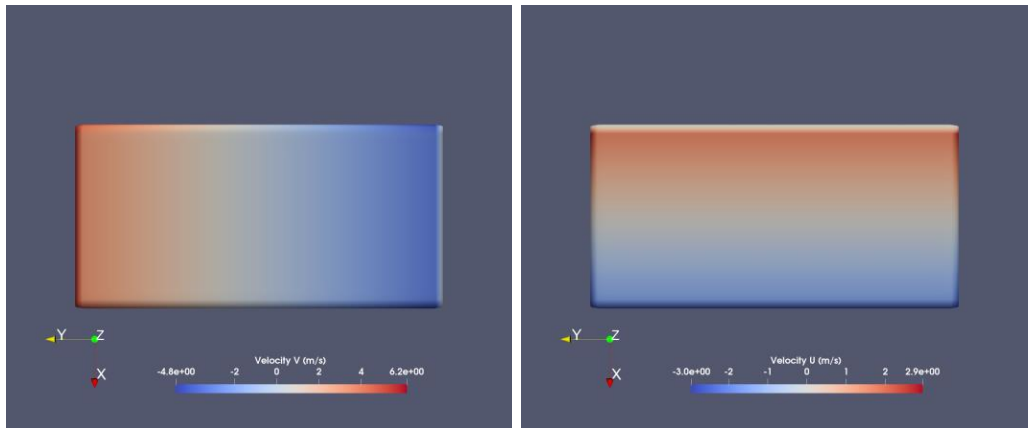

(a) Transverse velocity plotted on the contour of zero level set.

(b) Normal velocity plotted on the contour of zero level set.

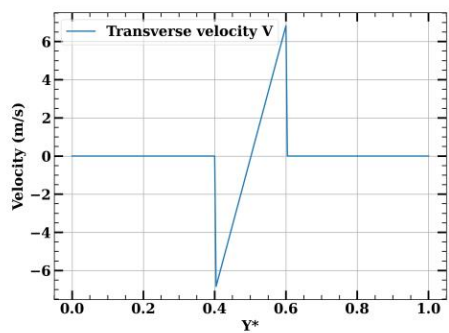
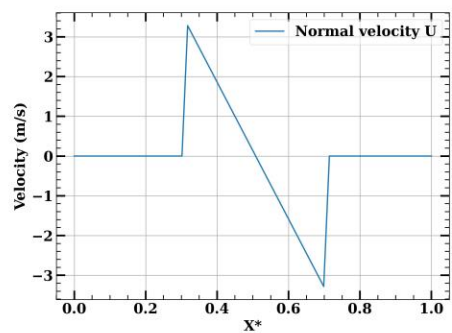

(c) Transverse velocity along y-axis at $x = 0$, $z = 0$.

(d) Normal velocity along x-axis at $y = 0$, $z = 0$.

Figure 2: Velocity profiles on the contour of zero level set. The value of zero is given to the gas phase outside of the zero level set. *dx*, $\theta = 5°$, $W = 80 m/s$. Dimensions are normalized by their respective axis ($L_y$ and $L_x$).

## 4. Overview of the DNS results

Fig. 3 provides a direct visualization of the DNS results obtained using the parameters specified in Table 1. The axial velocity, and consequently the liquid Weber number, was set sufficiently high to ensure the primary break-up occurs within the computational domain. From the front view, the liquid sheet is shown to stretch and diverge along the transverse direction. At any given axial location, the absolute transverse velocity is at a maximum at the sheet edges and decreases towards the center. The side view reveals that the sheet converges along the normal direction. Beyond a certain point, the outer contour of the spray begins to disperse in the normal direction. Reaching the state shown in Fig. 3 required an average of 40,000 iterations at the current mesh resolution, consuming approximately 10,000 CPU hours for each simulation.

Particular attention should be given to the early stage of the diverging jet before it fully develops into a stable liquid sheet. Fig. 3c shows a trim line that roughly separates two distinct regions. The central portion diverges normally and progresses smoothly in the axial direction, while the surfaces near the edges are significantly less regular. In Fig. 3d, an apparent triangular shape can be observed from the normal side. After exiting the injector with a rectangular shape, the liquid jet undergoes a transition to form a normal liquid sheet. The two sharp edges move toward the sheet's center in the normal direction and eventually merge. The conclusion of this process roughly marks the beginning of a stable liquid sheet formation. These initial features are related to the initialization of the distance function and the imposed velocity profiles, which could be refined in future work.

Nevertheless, these numerical artifacts gradually diminish and are smoothed out as the liquid sheet propagates downstream in the axial direction.

Using $\theta = 5°$ and the parameters listed in Table 1, different axial velocities $W$ (or liquid Weber numbers) were tested to evaluate their impact on spray characteristics. Fig. 22 presents the results across a wide range of axial velocities: 10 m/s (*We* = 570), 20 m/s (*We* = 2285), 40 m/s (*We* = 9183), 60 m/s (*We* = 20571), 80 m/s (*We* = 36571), and 100 m/s (*We* = 57142).

From *We* = 570 to *We* = 2285, an oscillating liquid stream expands to form a liquid sheet. Even at this low aspect ratio (sheet width to thickness), a key characteristic of flat fan sprays, the formation and growth of thick rims, is clearly observed. The substantial increase from $We$ = 2285 to $We$ = 9183 was chosen to induce aerodynamic break-up on the liquid sheet.

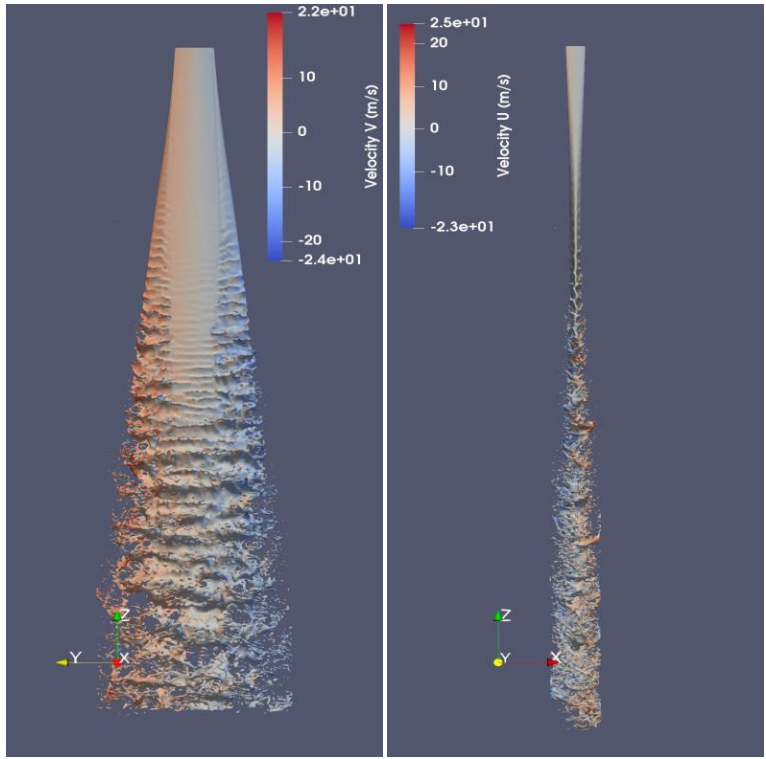

(a) Front view.      (b) Side view.

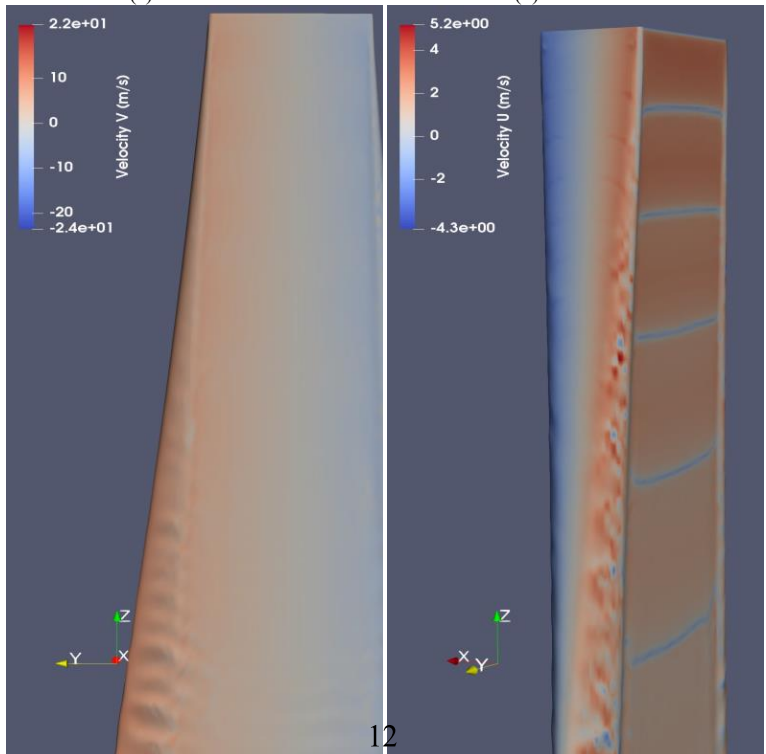

(c) Front view of the edge.      (d) Side view of the edge.

Figure 3: $\theta = 5°$, $W = 80 m/s$. Contours of the zero level set.

In the high Weber number regime, aerodynamic forces dominate the break-up process, manifested as K-H instabilities on the liquid sheet. Higher liquid Weber numbers result in earlier wave generation and more violent oscillations of the sheet in the normal direction.

The parameter $\theta$ plays a crucial role in controlling the divergence of the liquid sheet. In this work, $\theta$ was limited to $5°$, $10°$, and $15°$ to avoid numerical complications.

With $We$ = 20571 (axial velocity of 60 m/s) and other parameters from Table 1, results for the three $\theta$ values are shown in Fig. 23. From left to right, global thinning of the liquid sheet leads to stronger instabilities and earlier break-up, as reduced liquid inertia becomes more susceptible to external forces. Mesh resolution issues become apparent at higher $\theta$ values, where unphysical perforations accumulate at the sheet edge, promoting even earlier break-up compared to the sheet center.

Increasing the resolution yields more physically realistic liquid sheet patterns across various Weber numbers and $\theta$ values. To ensure the liquid sheet reaches the domain end at $\frac{dx}{2}$ resolution, each simulation ran for 2 days on 2048 CPUs, totaling approximately 100,000 CPU hours.

At $We$ = 2285 (Fig. 24), most instabilities visible at $dx$ resolution disappear at $\frac{dx}{2}$, suggesting that $dx$ resolution may be insufficient for detailed analysis of liquid structures. However, the liquid sheet remains bounded by two thick rims at both resolutions, indicating that $dx$ resolution is adequate for characterizing key global features, particularly rim evolution.

A comparison between the two resolutions at higher Weber numbers is provided in Fig. 25. Although the liquid sheet surface appears more irregular at higher resolution, certain characteristics persist: increased Weber numbers consistently lead to earlier break-up and enhanced instabilities. Equivalent results at high resolution across varying Weber numbers and $\theta$ values are shown in Fig. 26.

The current results are constrained by low aspect ratio limitations. The ratio of liquid sheet width to thickness before break-up ranges from approximately 10 to 50. In contrast, Sanadi (2022) reports ratios of hydraulic diameter to sheet thickness on the order of 50 to 100, before considering the ratio of sheet width to hydraulic diameter. The initialization method of the liquid sheet may also influence its atomization and break-up processes. Nevertheless, the present results provide a viable alternative for simulating this type of spray.

## 5. Velocity evolution

The transverse and normal velocity components are plotted along their respective $y$ and $x$ axes at different axial locations. The coordinates $x$, $y$, and $z$ are normalized by $l_x$, $l_y$, and $l_z$, respectively. Velocity values are sampled only in regions where liquid is present ($VOF > 0.9$). Consequently, discontinuities in the curves indicate either physical or unphysical break-up of the sheet.

In Fig. 4a, for $We = 2285$ and $\theta = 0$, the transverse velocity is zero at $z = 0$ by definition, indicating no diverging effect. This is confirmed by the relatively constant range of data along the transverse axis $y$ across different axial locations, showing a consistent liquid sheet width.

Increasing the value of $\theta$ results in a wider velocity distribution and a broader liquid sheet at the same axial location. Immediately downstream of the injection plane at $z^* = 0.125$, a sudden increase in transverse velocity is observed for all three non-zero $\theta$ configurations. This increase directly corresponds to the abnormal stretching process of the liquid sheet shown in Fig. 3.

A decrease in the absolute transverse velocity at the very edge of the liquid sheet is also visible in Fig. 4b, 4c, and 4d. This results from surface tension counteracting liquid inertia, pulling the edge liquid back toward the sheet center. This effect leads to the streamwise growth of the two thick rims that bound the liquid sheet, as seen in Fig. 24.

In Fig. 5, normal velocities are plotted at several axial locations. In contrast to the expansion along the transverse $y$-axis, the normal velocities act to contract the liquid sheet toward its center along the $x$-axis. Therefore, moving further downstream, a reduction in the data range along the $x$-axis, representing the decreasing sheet thickness, is both expected and observed.

The transverse and normal velocity profiles are shown again in Fig. 6 and Fig. 7, respectively, for $We = 20571$. At this high Weber number, the stretching effect persists across all configurations, as evidenced by the increased transverse velocity at the edge at $z^* = 0.125$. This confirms that the stretching effect originates primarily from the initialization process in the current configuration and is largely independent of the Weber number or the value of $\theta$.

The most notable difference from the lower Weber number case ($We = 2285$) is the reduced influence of the rim. At higher Weber numbers, surface tension can no longer sufficiently constrain the liquid inertia at the free edges,

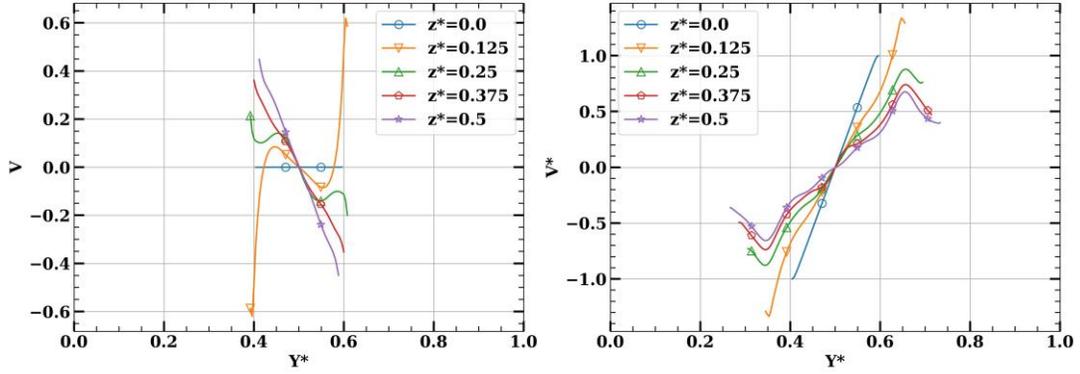

(a) θ = 0.

(b) θ = 5 ◦.

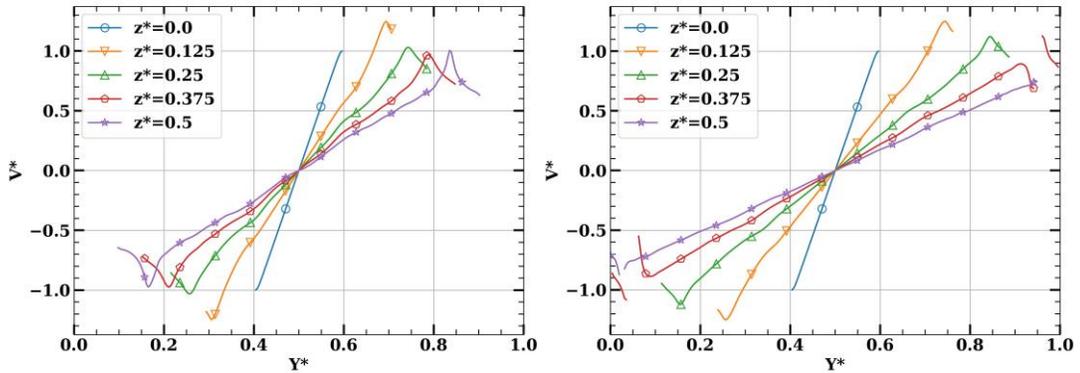

(c) θ = 10 ◦.

(d) θ = 15 ◦.

Figure 4: *We* = 2285, *dx*. Transverse velocity *V* plotted at *x* = 0 along the y-axis over several axial locations. When θ = 0', the velocities are normalized by the initial maximum transverse velocity at the edge ($WTan(\frac{\theta}{180°}\pi)$ for each configuration). The y-axis is normalized by $l_y$.

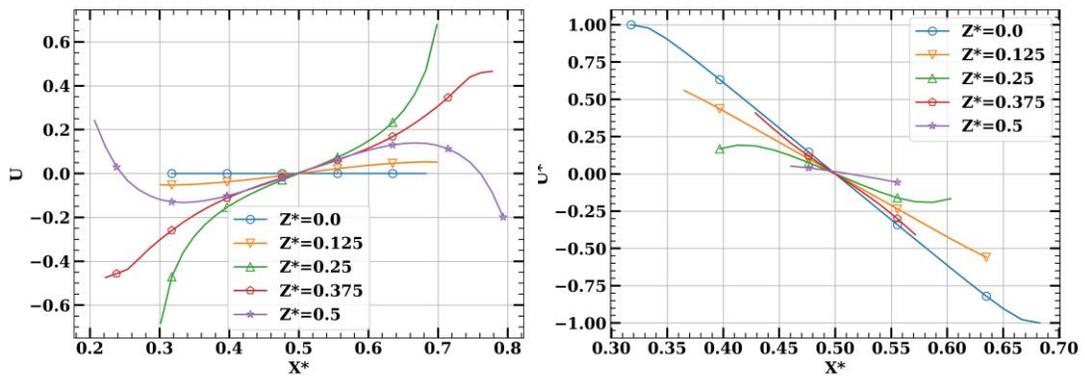

(a) θ = 0.

(b) θ = 5 ◦.

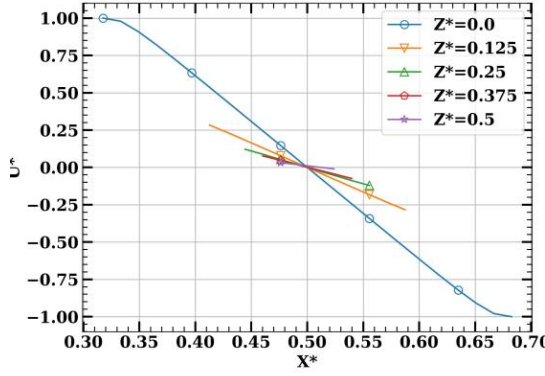
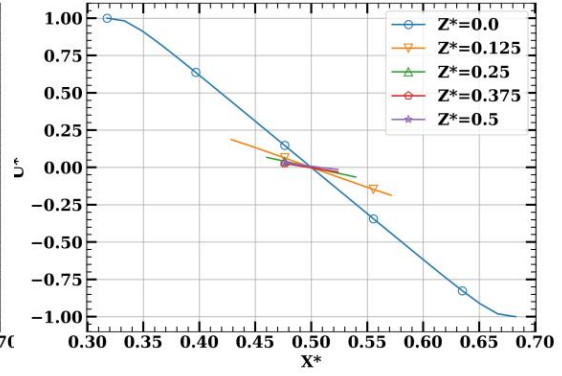

(c) $\theta = 10°$.

(d) $\theta = 15°$.

Figure 5: *We* = 2285, *dx*. Normal velocity *U* plotted at *y* = 0 along the x-axis over several axial locations. When $\theta = 0$, the velocities are normalized by the initial maximum normal velocity at the edge ($WTan(\frac{\theta}{180°}\pi)/2$ for each configuration). The x-axis is normalized by $l_x$.

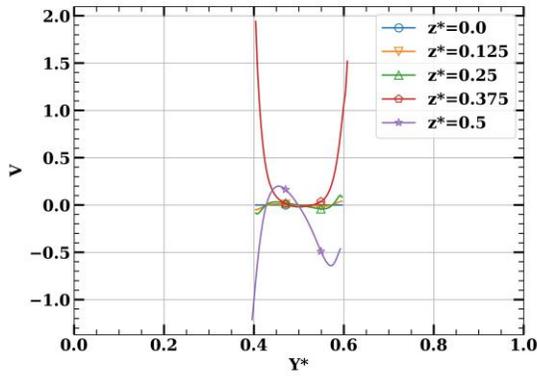
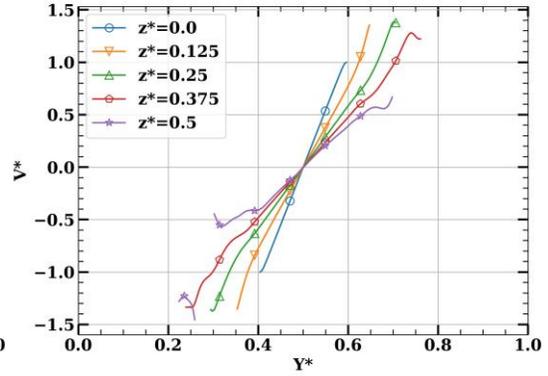

(a) $\theta = 0$.

(b) $\theta = 5°$.

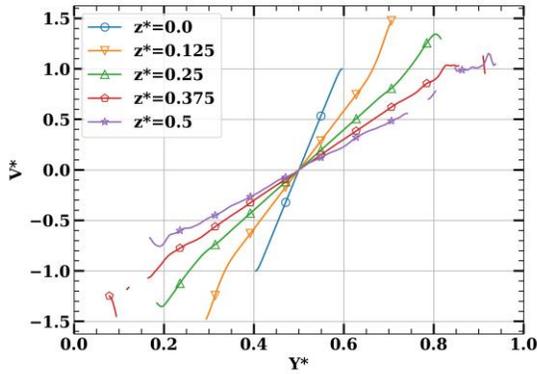
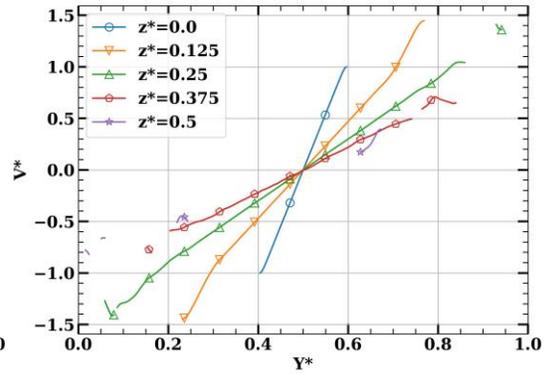

(c) $\theta = 10°$.

(d) $\theta = 15°$.

Figure 6: *We* = 20571, *dx*. Transverse velocity *V* plotted at *x* = 0 along the y-axis over several axial locations. When $\theta = 0°$, the velocities are normalized by the initial maximum transverse velocity at the edge ($WTan(\frac{\theta}{180°}\pi)$ for each configuration). The y-axis is normalized by $l_y$.

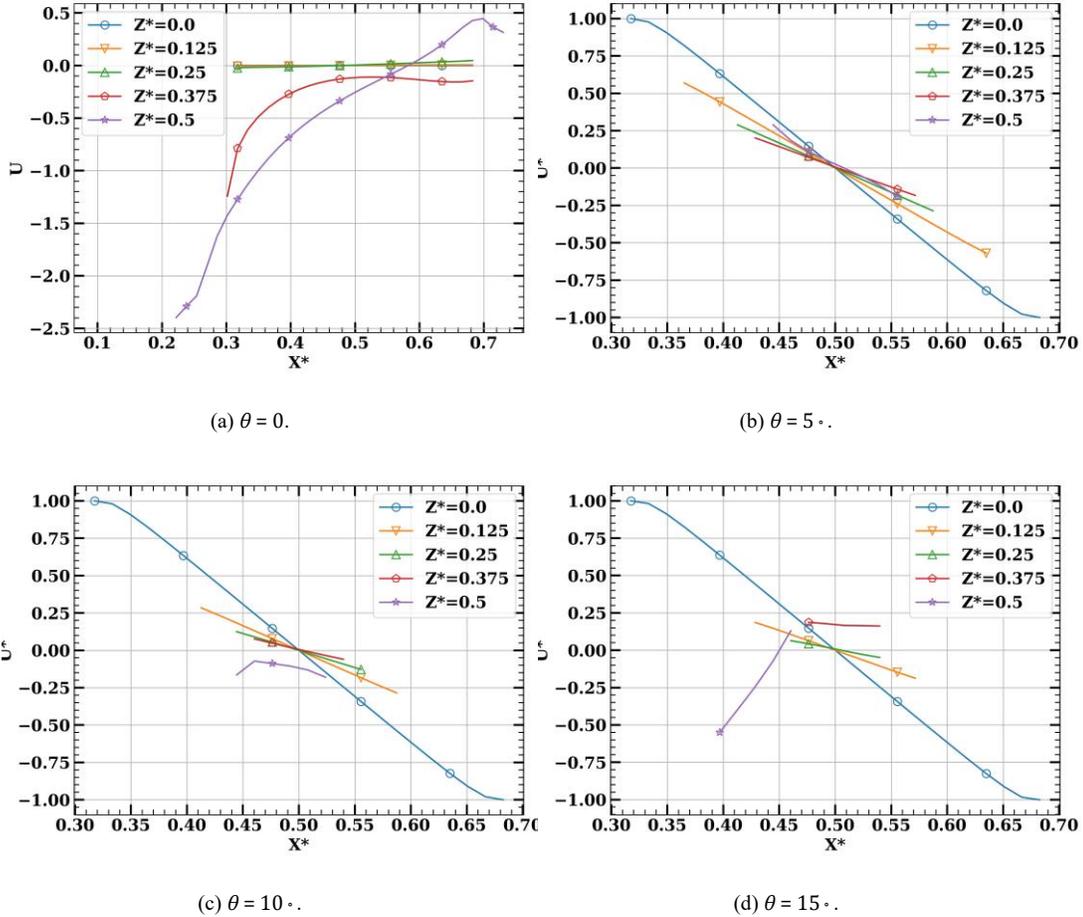

(a) $\theta = 0$.

(b) $\theta = 5°$.

(c) $\theta = 10°$.

(d) $\theta = 15°$.

Figure 7: *We* = 20571, *dx*. Normal velocity *U* plotted at *y* = 0 along the x-axis over several axial locations. When $\theta = 0°$, the velocities are normalized by the initial maximum normal velocity at the edge ($WTan(\frac{\theta}{180°}\pi)/2$ for each configuration). The x-axis is normalized by $l_x$.

and the transverse velocity increases almost linearly from the sheet center to its edge in the transverse direction.

Besides the effect on the edges, a higher Weber number also induces earlier break-up. More curve discontinuities appear in Fig. 6 compared to Fig. 4, where most curves remain continuous even at $z^* = 0.5$. A clear comparison at high resolution between different $\theta$ values in Fig. 26 illustrates the influence of the Weber number on determining break-up locations at the sheet edge, whether physical or not.

## 6. Transverse liquid sheet thickness

In addition to the velocity distributions, the liquid sheet thickness was measured along the transverse *y*-axis at several axial locations. The same configurations as before were used to better illustrate the influence of the Weber number on the liquid sheet morphology.

In Fig. 8 for *We* = 2285, as the liquid sheet propagates in the streamwise direction, the central thickness at *y* = 0 gradually decreases while the sheet width expands along the transverse *y*-axis.

Furthermore, the growth of the rims is observed more directly. The local maxima at the sheet edges correspond to the rim diameter, which increases downstream, while the local minima represent the thickness of the neck connecting the rim to the central liquid sheet. This behavior, clearly observed in Fig. 24, resembles the retraction of a free edge on a liquid sheet.

At *We* = 20571 in Fig. 9, the liquid sheet edges are significantly disturbed by aerodynamic forces. Although rim effects remain partially visible, the overall thickness distributions at the edges become much more irregular. Locations where the thickness drops to zero indicate the emergence of perforations and break-up events. However, these perforation are potentially under-resolved, suggesting that this break-up mode under the current configurations may be unphysical.

Sanadi (2022) employed time-gated optical diagnostics and classical backlight imaging techniques to capture detailed information of such liquid sheets. The existence of rims in flat fan sprays is effectively highlighted by intensity profile peaks in their backlight images, with experimental results showing clear correspondence with Fig. 8.

The farthest axial observation plane in Fig. 8 is positioned at the midpoint of the axial domain, corresponding to $20e_{inj}$. The maximum width at $\theta = 15°$ in Fig. 8d reaches approximately $10e_{inj}$, extending to both transverse

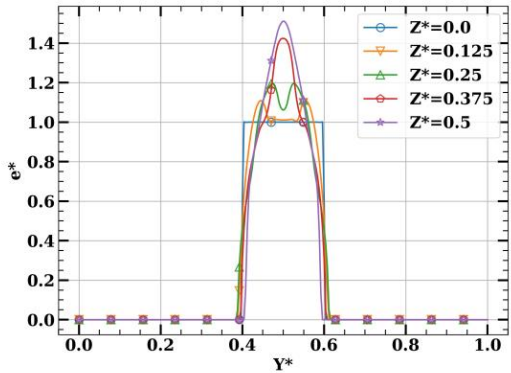

(a) $\theta = 0$.

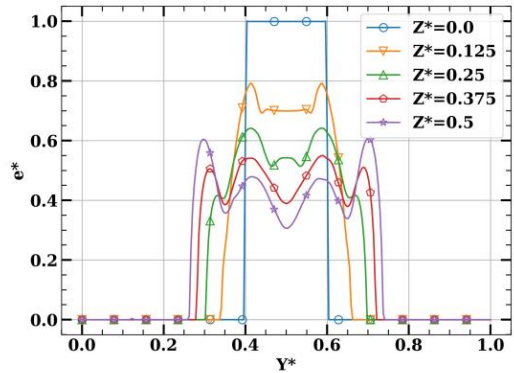

(b) $\theta = 5°$.

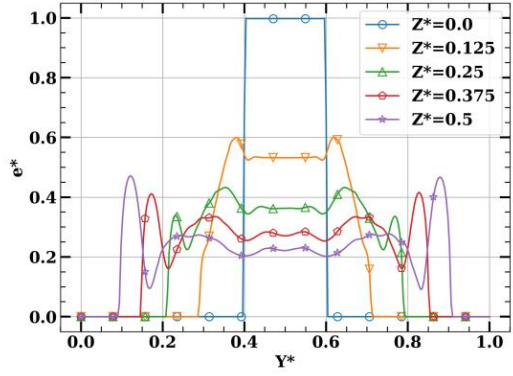

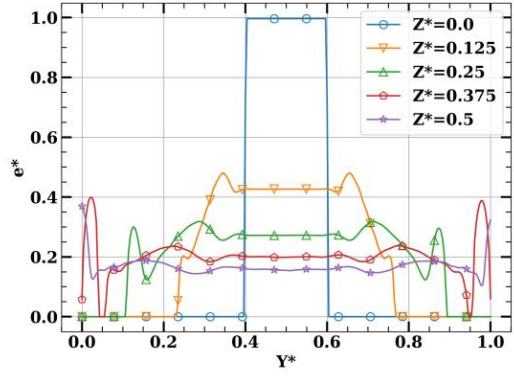

(c) θ = 10°.

(d) θ = 15°.

Figure 8: *We* = 2285, *dx*. The liquid sheet thickness at different transverse and axial locations is normalized by the initial width of the injector $e_{inj}$.

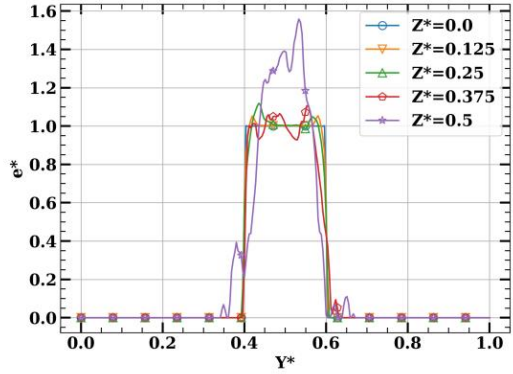

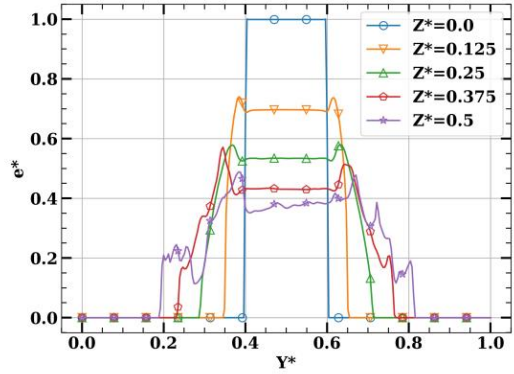

(a) θ = 0.

(b) θ = 5°.

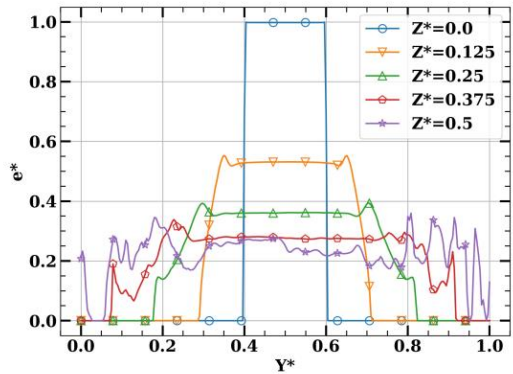

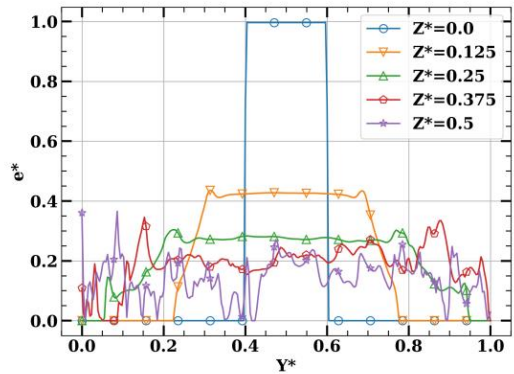

(c) θ = 10°.

(d) θ = 15°.

Figure 9: *We* = 20571, *dx*. The liquid sheet thickness at different transverse and axial locations is normalized by the initial width of the injector $e_{inj}$.

boundaries of the domain. In the work of Sanadi (2022), the observation window begins at $20D_h$ with a liquid sheet width of $20D_h$ where $D_h$ is the hydraulic diameter. This comparison indicates that the aspect ratio between the rim diameter and sheet thickness presented in Fig. 8 has great potential for further development.

Kashani et al. (2018) also provided insights into liquid sheet thickness distribution. Their transverse thickness measurements, taken close to the injector (2mm), reveal similar rim effects. The pronounced thickness variations observed near the sheet center and edge are attributed to flow disturbances that gradually diminish further downstream.

## 7. Evolution of the rim

As shown in Fig. 24, at low values of $\theta$ and Weber number, the liquid sheet remains bounded by two thick rims whose diameter evolves as the sheet propagates downstream from the injection plane. Fig. 10 presents zero level set contours at multiple axial locations, providing enhanced visualization of the rim evolution in space. As the liquid sheet stretches transversely, the rims grow through continuous liquid accumulation. These rims connect to the central liquid sheet via two progressively thinning necks. The adequate resolution of these necks is crucial for determining the physical accuracy of liquid cylinder or finger pinch-off. However, the large difference in size between the thinning necks and the expanding liquid sheet makes Direct Numerical Simulation of this spray configuration computationally prohibitive without advanced numerical techniques such as Adaptive Mesh Refinement (AMR).

In Fig. 11, one edge of the liquid sheet at $y = 0$ is used to collect the rim transverse velocity. The thickness of the liquid sheet center is collected at $y = 0$, the neck thickness and the rim thickness (diameter) are collected by finding the minima and maxima locally over a certain range of the y-axis.

In Fig. 11a and 11c, a peak of the rim velocity is observed at both resolutions. This probably signifies the end of the stretching process, a normal liquid sheet starts to form after getting rid of the effect of the initialization process. Only shortly after this stage, a rim and a neck begin to form when the thickness at the edge stops increasing monotonically towards the center. A crest and a trough are locally detected at low amplitudes as in Fig. 10. The rim diameter continues to grow as the neck thickness decreases and drops below the thickness of the liquid sheet center eventually. The noisy evolutions of the data in Fig. 11c and 11d can be largely reduced by increasing the mesh resolution. The abnormal peaks observed in Fig. 11a and 11b are due to the collision with floating droplets presented in Fig. 12.

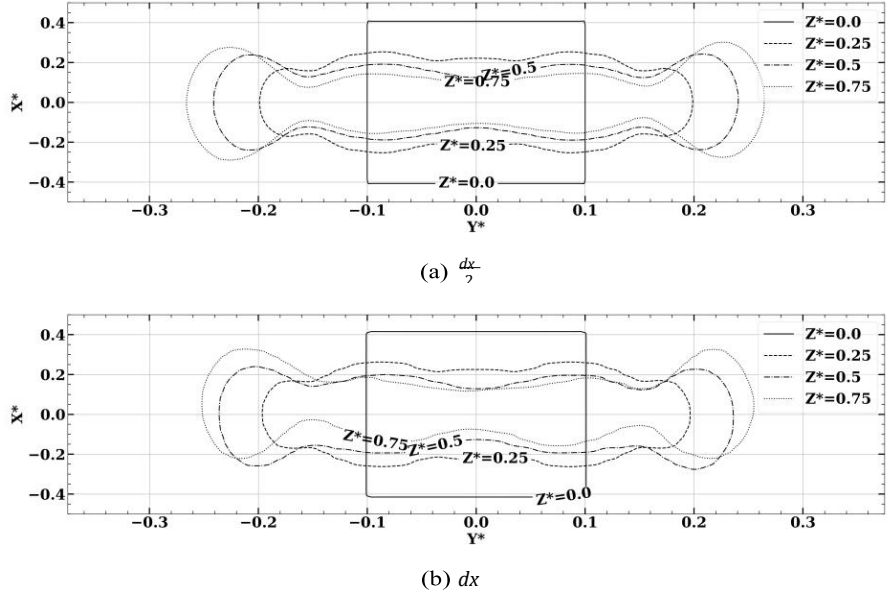

(a) $\frac{dx}{2}$

(b) $dx$

Figure 10: $\theta = 5°$, $We = 2285$. Contours of the cross-section of the liquid sheet at different axial locations. The x-axis and the y-axis are normalized by $l_x$ and $l_y$ respectively.

The rim transverse velocity reaches zero at both resolutions and at certain axial locations. At this point, the liquid sheet edge stops diverging under the effect of the surface tension while the liquid sheet center thickness continues to drop.

Taylor (1959) and Culick (1960) independently showed that the tip velocity of a capillary-driven retracting inviscid planar sheet reaches a steady value, now referred to as the Taylor-Culick velocity. This formula was obtained by balancing surface tension and inertia effects, assuming that the mass of the retracting sheet accumulates in a circular rim. It can also be used to characterize the retraction speed of the free edge of a liquid sheet (Deka et al., 2019). The Taylor-Culick velocity reads:

$$v = \sqrt{\frac{2\sigma}{\rho\tau}} \quad (6)$$

$v$ is the retraction speed of the liquid sheet, $\sigma$ is the surface tension of

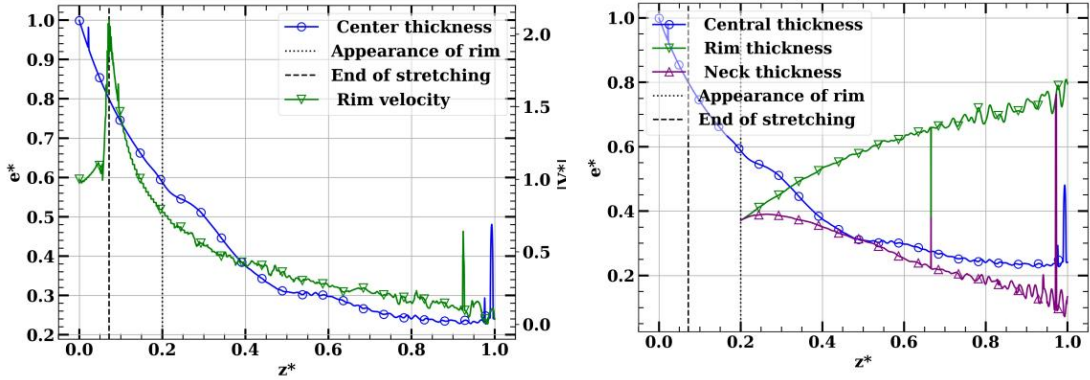

(a) $\frac{dx}{2}$. Variation of the rim transverse velocity (b) $\frac{dx}{2}$. Variation of the rim, neck and the liquid and the thickness of the liquid sheet center. sheet center thickness.

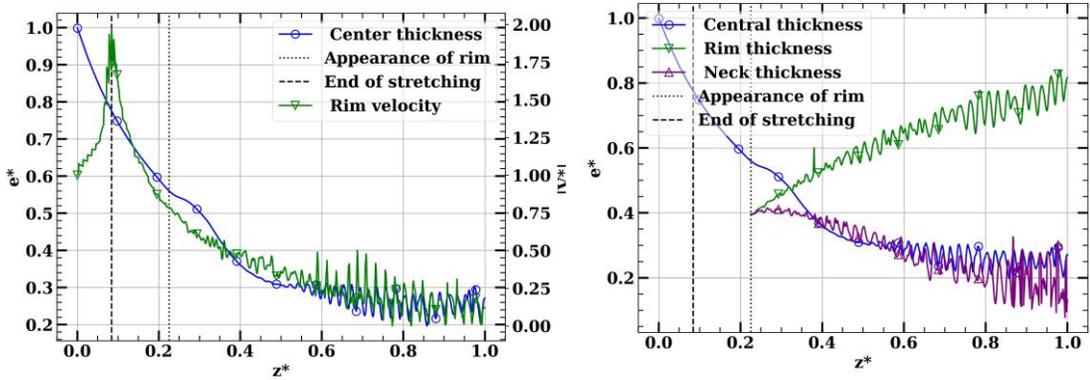

(c) *dx*. Variation of the rim transverse velocity (d) *dx*. Variation of the rim, neck and the liquid and the thickness of the liquid sheet center. sheet center thickness.

Figure 11: $\theta = 5°$, *We* = 2285 (*W* = 20*m/s*), Top row: $\frac{dx}{2}$. Bottom row: *dx*. Comparison between different thicknesses at different locations of the liquid sheet and rim transverse velocity. The thickness $e^*$ is normalized by the width of the injector $e_{inj}$, the rim transverse velocity is normalized by the initial maximum transverse velocity at the edge ($WTan(\frac{\theta}{180°}\pi)$).

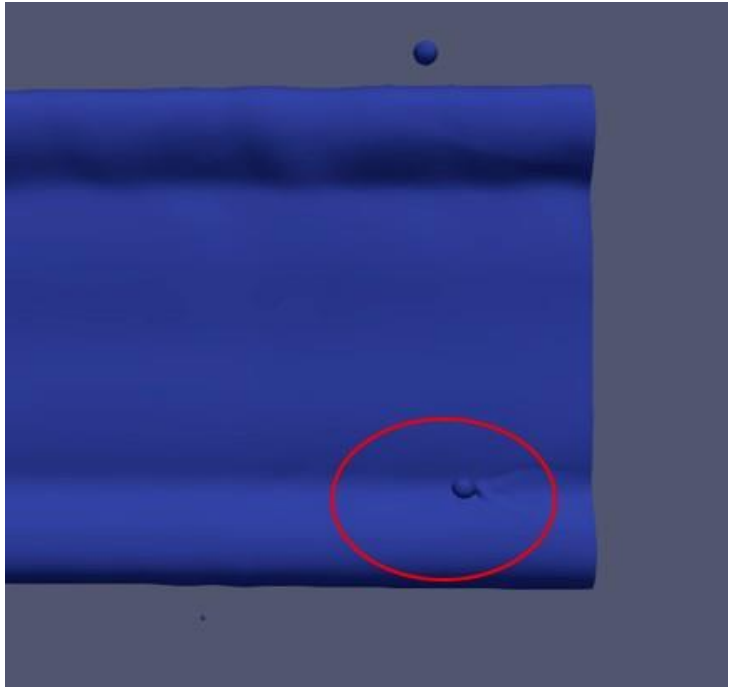

Figure 12: Floating droplets collide with the liquid sheet.

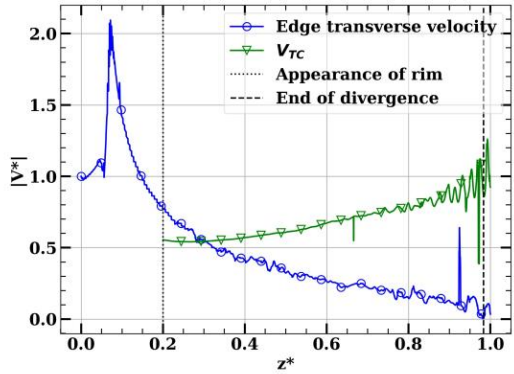

Figure 13: $\frac{dx}{2}$, $\theta = 5°$, $We = 2285$, Comparison between the rim transverse velocity and the Taylor-Culick velocity $V_{tc}$.

the liquid, and $\tau$ is the thickness of the sheet at the location of the hole or the rim.

Here, Eq. 6 can be introduced to compute the rim retraction velocity using the neck thickness in comparison with the actual rim transverse velocity. Fig. 13 shows the results of the comparison between these two velocities. It is worth mentioning that at the end of the divergence, where the rim transverse velocity reaches zero for the first time, $V_{tc}$ equals the normalized rim transverse velocity (0.75) at the point where the rim appears. Even though no publications are found focusing on the evolution of the rim of such a flat fan spray, the current comparison shows that at a low Weber number, the rim retraction speed of a free edge with Eq. 6 effectively accounts for the decrease of transverse velocity of the rim.

**8. Liquid sheet thickness estimation**

Dombrowski et al. (1960) first initialized the thickness measurement for flat fan spray. They carried out experimental and theoretical works and proved that the thickness is inversely proportional to the axial distance traveled by the liquid sheet. Following this idea, Sanadi (2022) performed experimental research on similar sprays to study the evolution of liquid sheet thickness. A geometrical model based on experimental parameters such as the opening angle of the nozzle has been created to estimate the thickness evolution. The estimated thickness was in excellent agreement with the measured thickness.

Based on the findings of previous works (Dombrowski et al., 1960; Sanadi, 2022), a simplified schematic is presented in Fig. 14 to match our current configurations. The liquid sheet is assumed to be part of a triangle. A virtual origin is proposed based on previous works and $Z_0$ is the axial distance from this origin to the injection plane 0. Plane 1 is set at an axial location after stretching to avoid potential bias. $L_1$, $e_1$, and $Z_1$ are the liquid sheet width, center thickness, and axial distance from the injection plane 0 to plane 1 and are all measurable. At plane $x$ where $Z_x$ is a random axial distance, the width and thickness $L_x$, $e_x$ need to be estimated and compared to our simulations.

From virtual origin to plane 1, with the assumption that the liquid sheet shown in Fig. 14 forms a triangle, $Z_0$ can be estimated using the following:

$$Tan(\theta) = \frac{\frac{L_1}{2}}{Z_0 + Z_1} \tag{7}$$

$$Z_0 = \frac{L_1}{2Tan(\theta)} - Z_1 \tag{8}$$

Using the same strategy, the liquid sheet width $L_x$ at a ramdom axial location $Z_x$ can be estimated using the previously calculated $Z_0$ and the measurable parameter set $L_1$, $e_1$, and $Z_1$:

$$Tan(\theta) = \frac{\frac{L_x}{2}}{Z_0 + Z_x} \tag{9}$$

$$L_x = 2Tan(\theta) * (Z_0 + Z_x) \tag{10}$$

Depending on different operating conditions, the liquid sheet edges carry different shapes, and a quantitative comparison between the estimated width $L_x$ and the measured one is questionable (e.g., case in Fig. 10). Nevertheless, the evolution of the cross-section follows a mass balance and its shape can be simplified as a rectangle of length $L_x$ and width $e_x$ at a random location $Z_x$. In this case, the estimated $L_x$ serves only to calculate the center thickness.

With the previous assumptions, the surface of the cross-section of the liquid sheet at plane x could be calculated using:

$$e_x L_x = e_1 L_1 \tag{11}$$

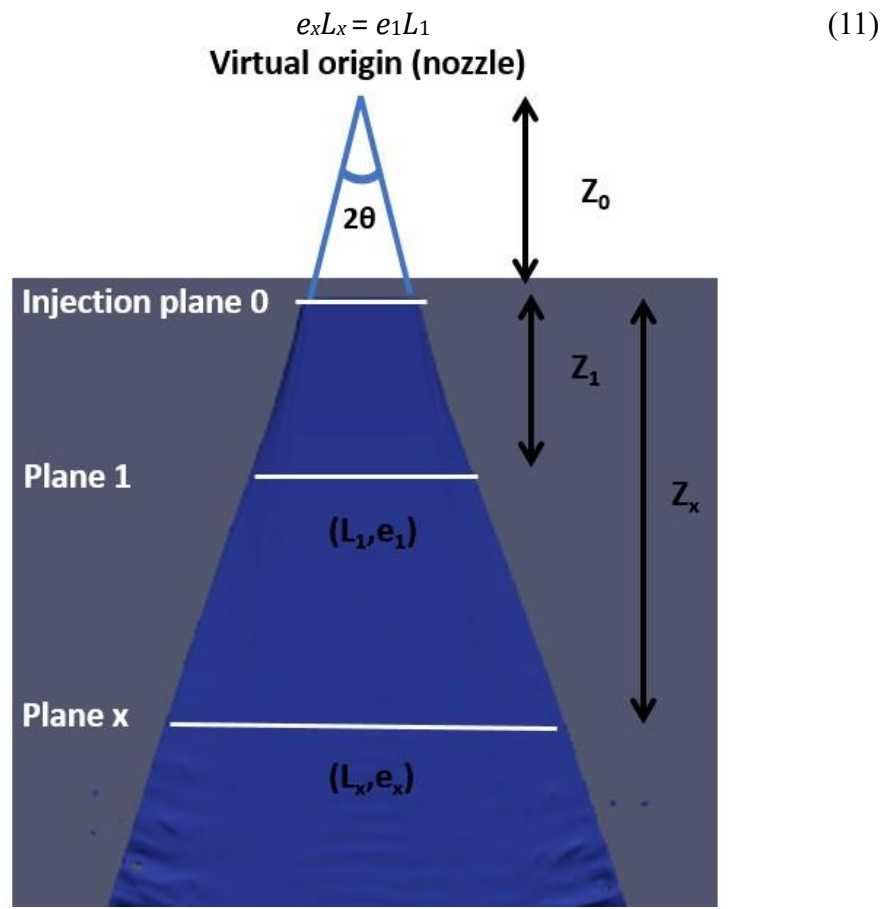

Figure 14: Schematic of the simplified geometry of the flat fan spray from the normal direction. L, e, and z are the liquid sheet width, center thickness, and the axial distance. The subscripts 0, 1, and $x$ denote different axial locations where previous parameters are measured/estimated.

Replacing $L_x$ with Eq. 10 and isolating $e_x$ yields:

$$e_x = \frac{e_1 L_1}{2 Tan(\theta) * (Z_0 + Z_x)} \tag{12}$$

Substituting $Z_0$ with Eq. 8 in Eq. 12 gives the following:

$$e_x = \frac{e_1 L_1}{L_1 + 2Tan(\theta) * (Z_x - Z_1)} \quad (13)$$

Which is the estimation of the liquid sheet center thickness based on known parameters. In Fig. 15, 12 configurations at various values of $\theta$ and $We$ have been used to test the accuracy. The plane 1 is set at $z^* = 0.125$ ($z_1 = 0.002$) to collect $L_1$ and $e_1$ for all configurations as an example. The liquid sheet center thickness is measured by summing the total volume over the whole x-axis using VOF at $Y = 0$ at each axial location, assuming that the liquid sheet is smooth without any ripples. The estimated liquid sheet center thickness shows very good agreement with the measured thickness from simulations.

The mean percentage error (MPE) can be computed for each one of the cases presented in Fig. 15 using the following equation:

$$MPE = \frac{\sum_{Z^*=0}^{Z^*=0.5} \left| \frac{e_{measured} - e_{estimation}}{e_{measured}} \right| * 100\%}{512} \quad (14)$$

Half the axial length contains 512 cells. The computed MPE is presented in Fig. 16. As $We$ and $\theta$ start to increase, the global liquid sheet tends to destabilize and flap at earlier stages, bringing more difficulties to our measurements.

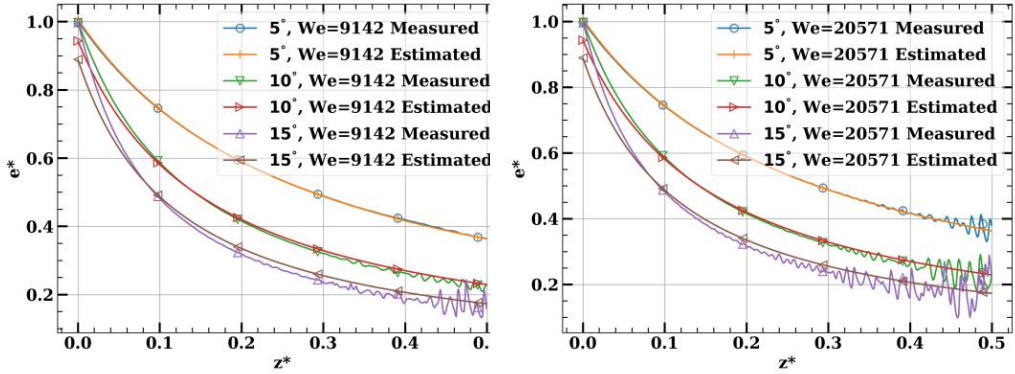

(a) $We$ = 9143  (b) $We$ = 20571

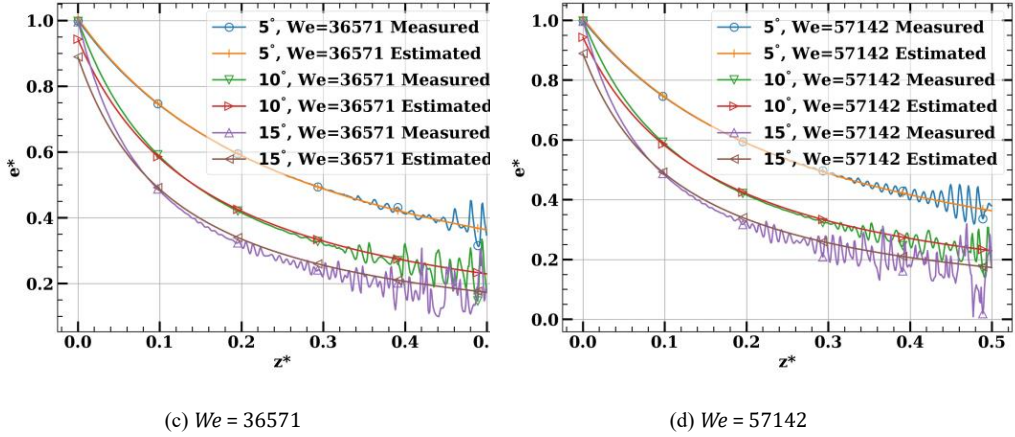

(c) *We* = 36571

(d) *We* = 57142

Figure 15: Comparison between the measured and estimated liquid sheet center thickness at various *θ*, *We* at resolution *dx*.

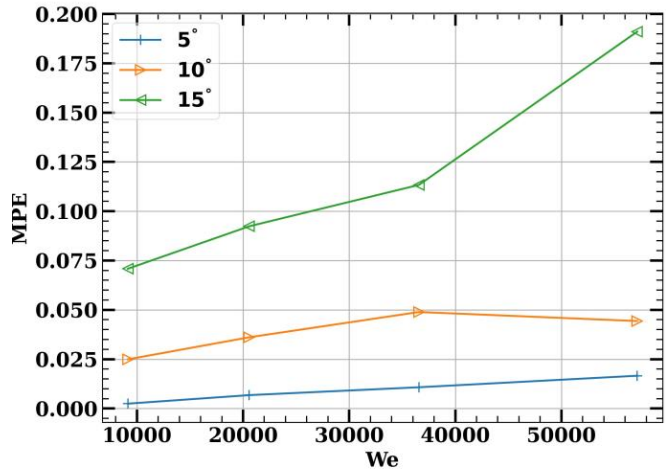

Figure 16: The mean percentage error is calculated for each configuration. The percentage error between measurement and estimation is collected at each axial location, and then a total of 512 values (over the first half of the axial Z-axis) is averaged.

## 9. Wavelength & amplitude

Clark and Dombrowski (1972) investigated the wavelength of maximum growth rate and the break-up length on attenuating sheets. Fig. 17 shows a comparison between a typical liquid sheet issued from a flat fan spray nozzle in experiments (Clark and Dombrowski, 1972) and one of our simulations. Two zones of wave growth have been identified in both situations. The waves that grow on the edge of the sheets are found to be generated much earlier near the injector. From the view of the image taken from experiments (Clark and Dombrowski, 1972), the instability triggered by the turbulence at the level of the nozzle propagates throughout the entire liquid sheet edge. From the view of the simulation, the velocity magnitude at the edge (3 components) is higher than the one in the center (velocities in the transverse and normal direction equal zero), especially at high values of $\theta$. This could also lead to stronger and earlier instabilities.

In the work of Dombrowski and Johns (1963) where particular attention is paid to the attenuating nature of the sheet, for an inviscid sheet, the dominant wave number is independent of the sheet thickness. However, for a viscous sheet, the maximum growth rate and the corresponding wave number increase as the sheet thickness decreases.

The sheet thickness at the edges is found to be much smaller than the center thickness in all of our configurations at high *We*. The wavelengths on these edges are also found to be much smaller than the ones in the center for both cases, which is in favor of the previous findings. However, the formation of these edges in current simulations is potentially influenced by the stretching process discussed earlier or even under-resolved.

Therefore, we draw our attention to the waves in the central region. In Fig. 18, windows are positioned over the same range of the z-axis for each one of the simulations. When increasing *We* from the left to the right, more pronounced irregularities are observed on the liquid sheet surface. Wave amplitudes are visibly higher and more detached liquid structures are generated at an early stage. When increasing the values of $\theta$ from top to bottom, the liquid sheet is thinner. A slight reduction of the wavelength is observed when taking *We* = 9143.

An attempt to recover the information on these waves is expected to qualitatively discover the links between the liquid sheet thickness, wavelength, and wave amplitude. As a step zero, the thickness of the half liquid sheet (a slice of VOF at $Y^* = 0$) is computed by summing the values of VOF at each axial location for each half liquid sheet in Fig. 18. In step one, the estimated thickness is computed with Eq. 13 for each configuration, a division by 2 gives the estimated half liquid sheet thickness. The estimated half liquid sheet thickness is then used to subtract the measured one in step zero. This aims to partially remove the effect of the attenuating thickness and to keep only the pure variation of the waves. In step two, the mean of the wave amplitude over the range of data in step one is subtracted from itself for better illustration.

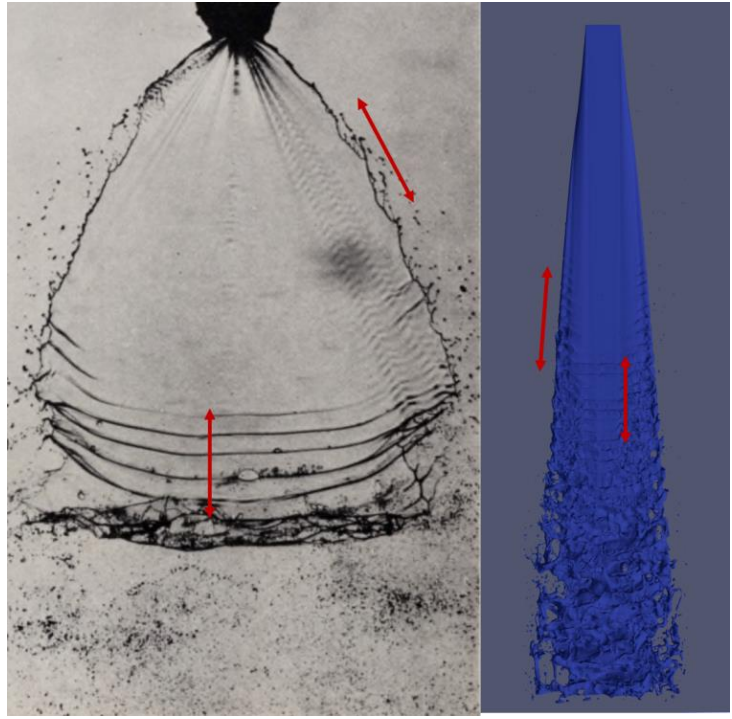

Figure 17: Left: image of an attenuating liquid sheet from the work of Clark and Dombrowski (1972) at *Re* = 10000 . Right: simulation result with $\theta$ = 5 °, *We* = 9143, fine mesh. In both cases, waves with different wavelengths and amplitudes are found in the liquid sheet center and at the edge. Red arrows indicate the regions where waves are observed.

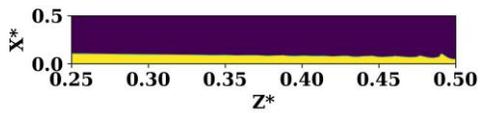
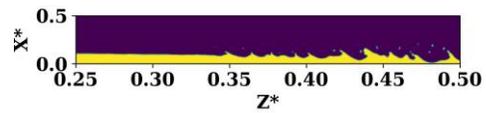

(a) $\theta$ = 5 °, *We* = 9143  (b) $\theta$ = 5 °, *We* = 57142

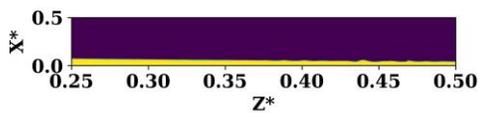
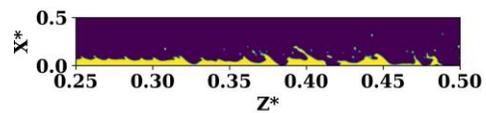

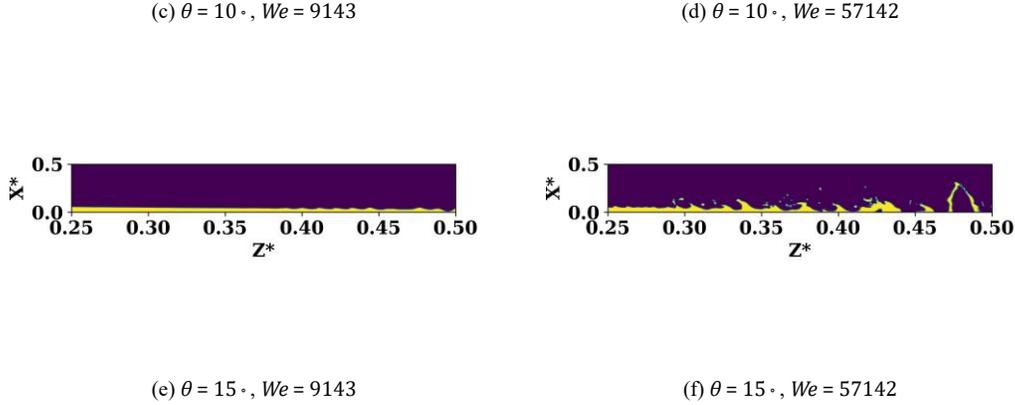

(c) $\theta = 10°$, We = 9143  (d) $\theta = 10°$, We = 57142

(e) $\theta = 15°$, We = 9143  (f) $\theta = 15°$, We = 57142

Figure 18: Fine mesh. Slice of the Volume-of-Fluid of the liquid sheet at $y = 0$ over $0.25 < Z^* < 0.5$. Only half ($X^* > 0$) is presented. Axes are normalized by their respective physical dimensions ($l_x$ or $l_z$).

The results at each step are presented in Fig. 19 for each slice of VOF presented in Fig. 18. The presented process is merely an initiative to reconstruct the signal. The crests are detected for each set of data from step two to compute the mean wavelength and amplitude.

Fig. 20a shows the computed mean wavelength for different configurations at two resolutions. When using coarse mesh, passing the data through the previous process does not provide comprehensive results. With fine mesh only, the previous process shows initial patterns varying We and $\theta$. The increase of the value of $\theta$ or the decrease of liquid sheet thickness leads to a slight decrease in wavelength. This is in correspondence with the findings of Dombrowski and Johns (1963). The increase of We has similar effects but is less pronounced due to the noisy nature of the data collected this way.

The Squire wavelength is originally proposed in the work of Squire (1953), where the stability of a thin liquid sheet moving in still air is studied. It is found that the instabilities occur when $W = \frac{T}{\rho_1 U^2 h} < 1$ where $T$ is the surface tension of the liquid, $\rho_1$ is the liquid density, $U$ is the liquid velocity and h is the half thickness of the liquid sheet. The wavelength of the maximum growth rate, for $W << 1$, is $\lambda = \frac{4\pi T}{\rho_2 U^2}$ where $\rho_2$ is the density of the surrounding gas.

Villermaux and Clanet (2002) derived the following expression for circular sheets:

$$\lambda = \frac{10\pi T}{\rho_a U^2}$$ (15)

Where $\rho_a$ is the air density. Eq. 15 is used (Gaillard et al., 2022) to compare to the data obtained from the flat fan spray of water. The experimental

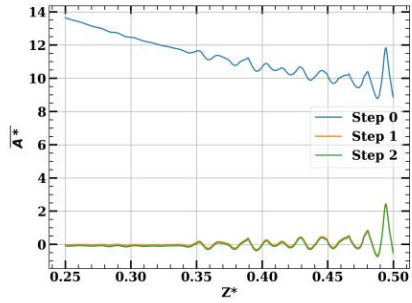

(a) $\theta = 5°$, We = 9143

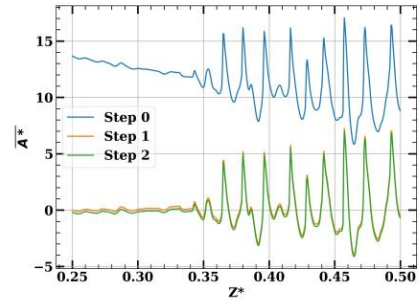

(b) $\theta = 5°$, We = 57142

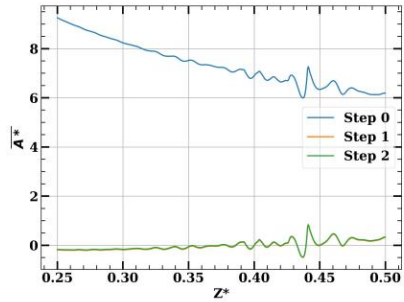

(c) $\theta = 10°$, We = 9143

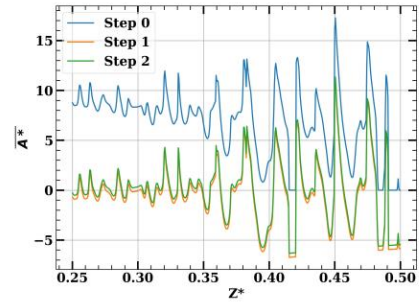

(d) $\theta = 10°$, We = 57142

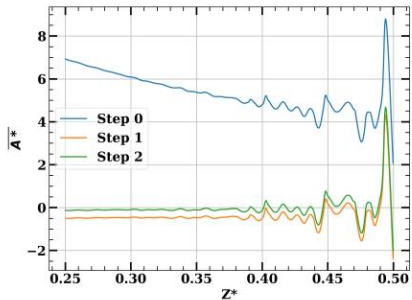

(e) $\theta = 15°$, We = 9143

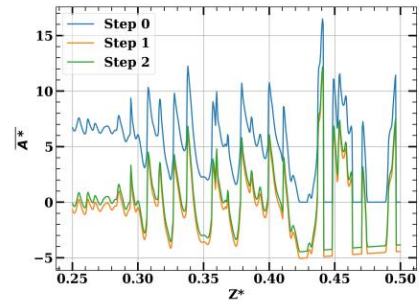

(f) $\theta = 15°$, We = 57142

Figure 19: Step zero: half liquid sheet thickness computed from Volume-of-Fluid. Step one: Estimated half liquid sheet thickness subtracted from step zero. Step two: Mean of the distribution from step one subtracted. The amplitudes are normalized by *dx* (fine mesh).

data are underpredicted by a factor of 2 at high velocities ($U = 30 m/s$ or *We* around 10000), this underprediction tends to grow as *We* is further increased.

Eq. 15 is used in Fig. 20a to compare to the computed wavelength described earlier. This comparison serves only as an indicator of magnitude to validate our observations of the simulations at minimum cost.

On the other hand, in Fig. 20b, earlier waves are generated and their amplitudes are much higher when increasing *We* compared to the effect varying $\theta$. In both cases, only fine mesh provides qualitative correlations.

The proposed process does not seek to provide a quantitative correlation between the liquid sheet thickness and the wavelength, especially at current mesh resolutions, and is open for future improvements. The slices of the liquid sheets in Fig. 18 are potentially unsuitable for such an analysis due to their irregularities and ruptures. Nevertheless, given the limited time and at an early phase of research on current configurations, it is meaningful to provide a comprehensive and qualitative comparison between the liquid sheet thickness and the wavelength. This comparison serves to validate our simulations based on existing findings.

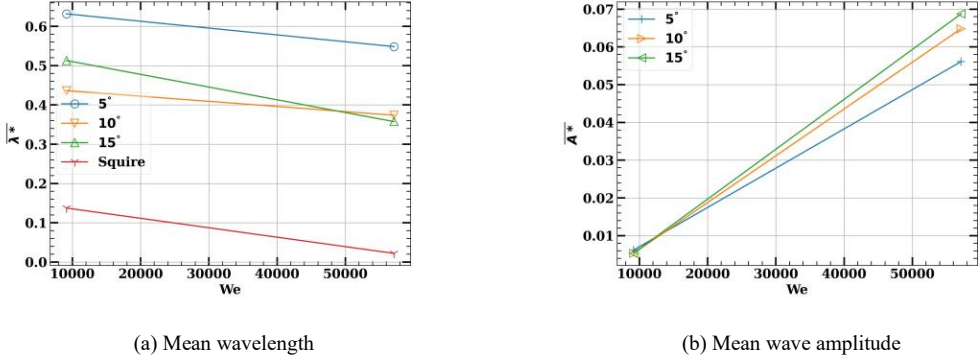

(a) Mean wavelength  (b) Mean wave amplitude

Figure 20: Mean wavelength and amplitude normalized by the width of the injector $e_{inj}$ for each set of the configuration.

## 10. Break-up length estimation

Last but not least, the investigation of the break-up length is of great importance in the atomization process of spray. Kooij et al. (2018) provided a general scaling of the break-up length based on empirical formulas and instability analysis. Their theories are further tested and validated with the experimental data from Gaillard et al. (2022) on fan spray nozzles.

In our current configurations of the flat fan spray, the opening angles of the spray are modeled by varying the values of $\theta$. Further tests on the estimation of the sheet thickness rely on proper geometrical modeling and show promising results. However, the derivations of Kooij are based on experimental configurations such as the nozzle's dimensions. Given the fact that the break-up of a standard liquid sheet has been extensively investigated, which is not the purpose of the current work, a different formulation is expected to suit the current framework (an attenuating liquid sheet or flat fan spray) incorporating the existing parameters.

Senecal et al. (1999) performed instability analysis on both parallel-sided and attenuating sheets. For a parallel-sided sheet where the half thickness $h$ doesn't depend on the radial distance, the following expression is proposed to predict the breakup length:

$$L = \frac{V}{\Omega_s} \ln \frac{\eta_b}{\eta_0} \tag{16}$$

$\eta_0$ is the initial amplitude of aerodynamic waves on liquid sheets and $\eta_b$ is the amplitude of aerodynamic waves at break-up length. $\ln \frac{\eta_b}{\eta_0}$ is given the value 12 (Dombrowski and Hooper, 1962) and $\Omega_s$ is the maximum growth rate of aerodynamic waves. However, for an attenuating liquid sheet where the half thickness $h$ is constantly changing as a function of the radial distance, the following formulation is proposed for the long wave mode:

$$L = V(3\ln\frac{\eta_b}{\eta_0})^{2/3}(\frac{J\sigma}{(\rho_g/\rho_l)^2 U^4 \sigma_l})^{1/3} \qquad (17)$$

J is a constant based on the inverse relation between the attenuating liquid sheet thickness and the distance traveled. $h = J/t$ is used in the work of Senecal et al. (1999) where t is the time traveled by the liquid sheet once leaving the nozzle. $2h = Ut^K$ is used in the work of Clark and Dombrowski (1972) where K is a constant and U is the absolute velocity of the liquid. The detailed derivation of these results is beyond the scope of the current work and can be found in the work of Senecal et al. (1999).

Eq. 16 (independent of sheet thickness) is used for the short wave regime (*We* > 27/16), which is our case. However, judging from Fig. 26, with the increase of $\theta$ or the decrease of sheet thickness, instabilities grow earlier on the liquid sheets and trigger eventually an earlier break-up. The liquid sheet surface also seems to be sufficiently resolved. These observations show that, in current configurations, a formulation incorporating the liquid sheet thickness is needed.

With the purpose mentioned earlier, we turn our attention to Eq. 17 even if it is only suitable for the long wave regime. The difference is that it requires explicit calculation of a thickness factor J, which matches well within the current frameworks.

Following the definitions in the work of Senecal et al. (1999), assuming the attenuating liquid sheet with thickness $2h$ or $e$ moving downstream with constant axial velocity *W* at time *t*, we have:

$$h = \frac{J}{t} \qquad (18)$$

Time *t* could be substituted with the constant axial velocity *W* and the axial distance traveled. In the experiment, this distance originates from the nozzle. In current configurations, this distance equals $Z_0 + Z_x$ where $Z_0$ is the calculable distance from the virtual origin to the plane of injection, $Z_x$ is the measurable distance from the plane of injection to any random plane of interest. Therefore:

$$h = \frac{JW}{Z_x + Z_0} \tag{19}$$

Replacing half thickness $h$ with thickness $e_x$ gives:

$$e_x = \frac{2JW}{Z_x + Z_0} \tag{20}$$

The liquid sheet center thickness is measurable information at any axial location ($e_1$ at $Z_1$ for instance), $Z_0$ can be calculated with Eq. 8. Eq. 20 becomes:

$$e_1 = \frac{2JW}{Z_1 + \frac{L_1}{2Tan(\theta)} - Z_1} \tag{21}$$

Isolating J gives:

$$J = \frac{e_1 L_1}{4W Tan(\theta)} \tag{22}$$

The value of J will then be calculated for each configuration varying axial velocities and values of $\theta$. $\ln \frac{\eta_b}{\eta_0}$ is given the value 12 (Weber, 1931). Clark and Dombrowski (1972) further correlated this constant to the Reynolds number and gave the value of 12 for $Re > 9000$ and 50 otherwise. The value of 12 is used to predict the mean drop size (Senecal et al., 1999; Dombrowski and Hooper, 1962) and to predict the break-up length (Asgarian, 2020) and shows good agreement with their simulation or experimental results.

Sarchami et al. (2009) investigated the initial perturbation amplitude of liquid sheets. They applied the correlation in the work of Inamura et al. (2004) to a wide range of experimental data (fan sprays (Dombrowski and Johns, 1963), splash-plate atomizers (Inamura et al., 2004; Ahmed et al., 2008)) to calculate the droplet diameter. However, their results show that each of the equations proposed (Dombrowski and Johns, 1963; Inamura et al., 2004; Ahmed et al., 2008) is only valid within its own specific range of operation. The existence of a wide range of correction factors in different correlations indicates that $\ln \frac{\eta_b}{\eta_0}$ may depend on liquid properties and actual flow conditions.

To tackle this problem, Sarchami et al. (2009) proposed the following break-up parameter based on fluid physical properties, nozzle diameter, and injection velocity in dimensionless forms:

$$\ln \frac{\eta_b}{\eta_0} = Re^{0.07} We^{0.37} \qquad (23)$$

Eq. 23 covers a wide range of *We* (from $10^4$ to $4*10^4$) and gives a wide range of values (from 40 to 200), which is one order of magnitude bigger than 12. Although no further publications are found to give more details based on the work of Sarchami et al. (2009), from the types of configuration (flat fan spray or splash-plate nozzles), range of high *We*, to a break-up mainly due to aerodynamic forces (instead of perforations or rim-driven break-up), Eq. 23 is a nearly perfect match for our operating conditions regardless of its precision level and empirical nature. Thus to this, substituting $\ln \frac{\eta_b}{\eta_0}$ with Eq. 23 in Eq. 17 gives the final break-up length estimation for an attenuating liquid sheet:

$$L_b = W(3Re^{0.07}We^{0.37})^{2/3}\left(\frac{J\sigma}{(\rho_g/\rho_l)^2 U^4 \sigma_l}\right)^{1/3} \qquad (24)$$

In Fig. 21, the break-up lengths measured from simulations under different $\theta$ and *We* are presented. A break-up is identified as long as a visible hole or gap is formed on the central region of the liquid sheet ($-L_{inj}/2 < Y < L_{inj}/2$). The liquid volume fractions or the evolution of $D_{32}$ cannot be used as such a criterion. As the liquid sheet stretches in the transverse direction *y*, its thickness decreases in the normal direction. As a consequence, the same amount of volume is conserved at different axial locations, and the $D_{32}$ does not exhibit a sudden decrease due to the continuously decreasing thickness (gradual transition of liquid volume-surface ratio).

Due to these issues, the measurement of break-up lengths with the previous method has a maximum error of ±10% of $L_z$ as a result of rough approximations. The predictions are computed using Eq. 24 varying several parameters for each of the configurations. Both predictions and measurements show the same tendency:

Both the increase of *We* or the thinning of the liquid sheet (the increase of *θ*) renders the sheet more unstable under the effect of aerodynamic forces and consequently leads to an earlier break-up.

Furthermore, different values are given to $\ln \frac{\eta_b}{\eta_0}$ based on either Eq. 23 or fixed values (12 (Senecal et al., 1999; Dombrowski and Hooper, 1962; Asgarian, 2020), 12 or 50 (Clark and Dombrowski, 1972) based on *Re*). Only Eq. 23 shows reasonable correlations between the break-up length and the Weber number, the break-up lengths computed with fixed values do not follow such a rule. Even after considering such a wide range of error (20% of $L_z$), in Fig. 26, the break-up lengths decrease visibly within the computational domain while increasing the Weber number, fixed values apparently go against this observation.

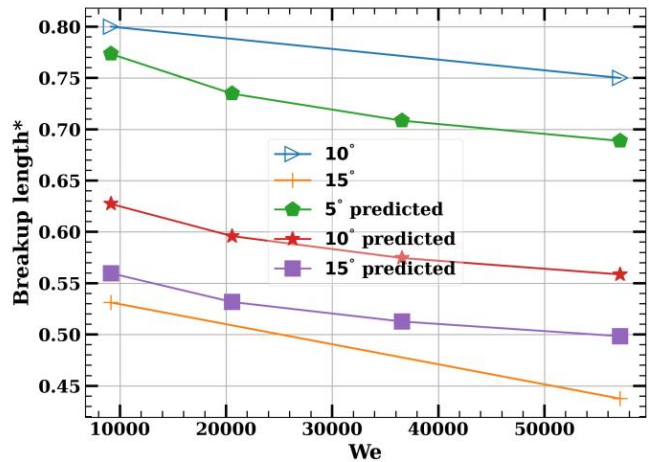

Figure 21: All lengths are normalized by $L_z$. Measured break-up lengths from simulations with fine mesh are shown (if a break-up is observed) in comparison with the predictions obtained from Eq. 24. An error of ±10% exists for all sets of measured lengths. No physical break-up is observed for *θ* = 5°.

## 11. Conclusion

In current works, direct numerical simulations of flat fan spray are carried out. Special transverse and normal velocity profiles are found to be crucial to recreate the feature of divergence of such a liquid sheet. The geometrical patterns of the liquid sheet can be effectively controlled by varying the value of the opening angle. This is potentially comparable to modifying the pressure to produce different flat fan sprays in experiments.

The behavior of the liquid sheet is investigated through several aspects. At a low Weber number, the surface tension eventually counteract the effect of centripetal forces and no break-up is observed. The transverse velocity at the edge is found to decrease slightly with respect to what has been applied initially. The fact that the liquid sheet is eventually bounded by two growing thick rims correspond to the findings in the literature. As the liquid sheet travels further downstream, it stops expanding in the transverse direction up to a certain point, To illustrate this point, the Taylor-Culick velocity is found in the literature and can be defined as the retraction speed of the free edge of a liquid sheet based on the liquid sheet thickness. Such a velocity is computed and is compared closely with the rim transverse velocity from the location where a rim starts to form till the end of divergence where the rim transverse velocity first equals zero. At the end of the divergence, the Taylor-Culick velocity is found to equal the rim transverse velocity when a rim starts to form. This is an indication that the growth of the rim and the decrease of the rim transverse velocity are physical and are not scenarios generated by the initial conditions like the abnormal stretching process.

At much higher Weber numbers, the surface tension fails to constrain the liquid inertia. The previous rim is not observed and the liquid sheet edge is shattered by strong aerodynamic instabilities. The transverse liquid sheet thickness distributions are less regular compared to the case with a low Weber number. In general, when the Weber number is the lowest, the liquid stream is formed, and a slight increase of the Weber number leads to a regular growth of the rim that can be explained by the balance between the surface tension and the liquid inetia. Further increasing the Weber number results in violent flappings and much earlier disintegrations. These are reasonable proofs that the current configuration is capable of reproducing the most important geometrical feature of such a liquid sheet after comparisons with different experimental or simulation results in the literature.

An attempt to estimate the central liquid sheet thickness axially can also be made based on a simplified geometry of the flat fan spray proposed in the literature. The estimated liquid sheet thickness using only initial conditions and information near the injection plane agrees very well with the measured one from simulations at various conditions. This further supports the fidelity of current set-ups trying to reproduce a realistic flat fan spray.

Utilizing the proposed simplified geometry, wave formation on the flat fan spray was captured across a range of opening angles and Weber numbers. The results indicate that an increase in liquid sheet velocity, and consequently the Weber number, produces the expected growth in wave amplitudes and leads to a more irregular and chaotic sheet surface. Furthermore, a larger opening angle or a reduction in sheet thickness appears to promote a decrease in the observed wavelength. These findings are in partial agreement with the limited experimental data available for this and similar configurations.

Finally, an effort was made to predict the break-up length of the flat fan spray, which is representative of liquid sheets with attenuating thickness. Established empirical and theoretical formulas from the literature, derived from experimental measurements and linear instability analysis, were employed, incorporating key parameters from the current configurations such as sheet thickness and axial velocity. The predicted break-up lengths were found to be of the same order of magnitude as those measured from the simulation results. Both datasets exhibit consistent trends, with the breakup length decreasing slightly with increasing opening angle or axial velocity. These correlations can serve as valuable indicators for the more systematic numerical prediction of break-up length in such sprays.

However, other challenges still exist. Only one of the disintegration modes is observed. The typical rim driven break-up is not observed in any scenarios and requires further implementation of other numerical tools such as the AMR (Adaptive Mesh Refinement). According to the literature, the hole break-up also requires a proper modeling of the nozzle geometry using different mesh strategies.

To conclude, to authors' best knowledge, the direct numerical simulations of flat fan spray haven't been performed in the literature, limited publications address directly to the formation of the rim, liquid sheet thickness evolutions and many other typical features. Therefore, the current work attempts to provide some insights regarding the characterization of the flat fan spray numerically and has the potential to be further investigated.

**Acknowledgments**

Authors from Université de Rouen Normandie thankfully acknowledge the computer resources at CRIANN (Centre Régional Informatique et dApplications Numériques de Normandie).

**Conflict of interest**

The authors have no conflicts of interest to disclose.

**Data availability**

The data that support the findings of this study are available from the corresponding author upon reasonable request.

**Appendix**

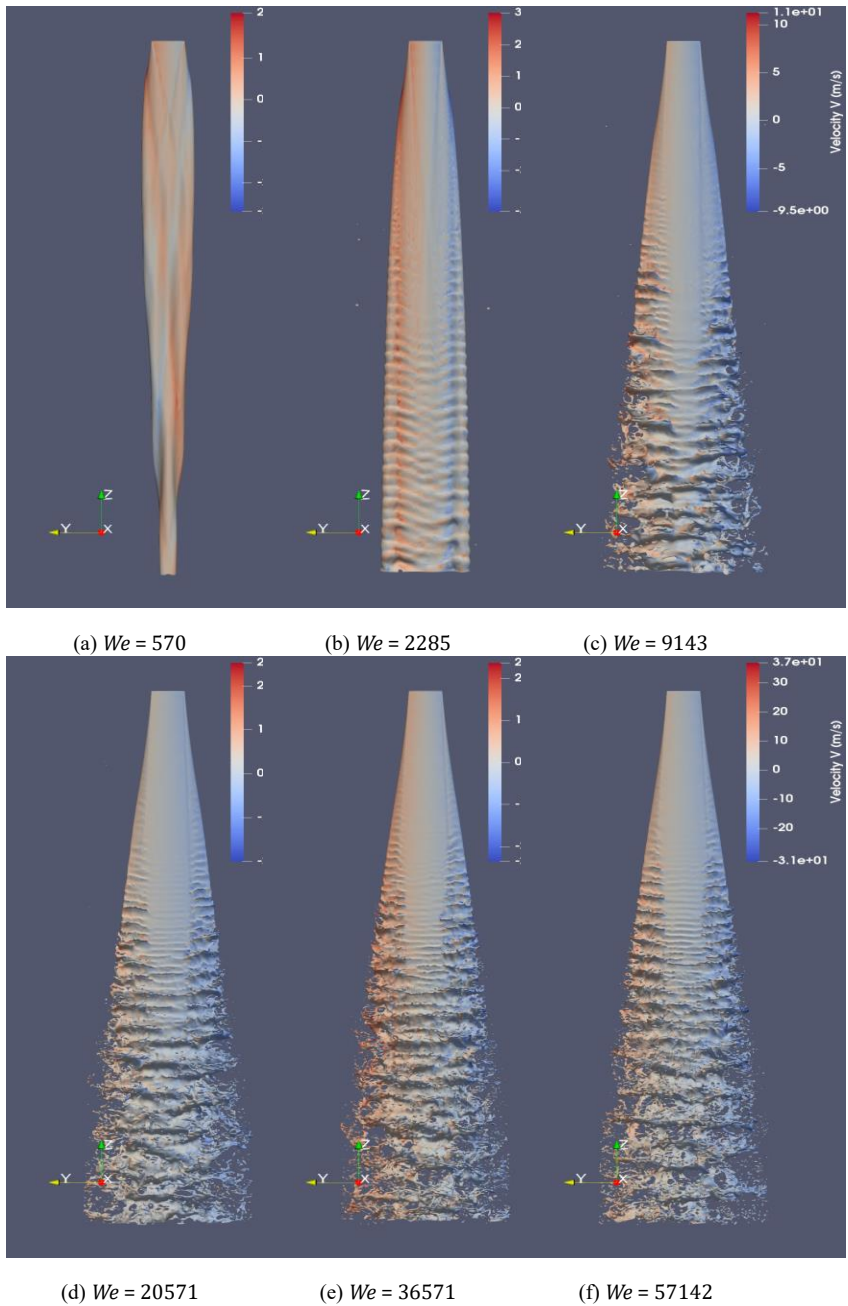

(a) *We* = 570  (b) *We* = 2285  (c) *We* = 9143

(d) *We* = 20571  (e) *We* = 36571  (f) *We* = 57142

Figure 22: *dx*, $\theta$ = 5°. Contours of zero level set at various Weber numbers.

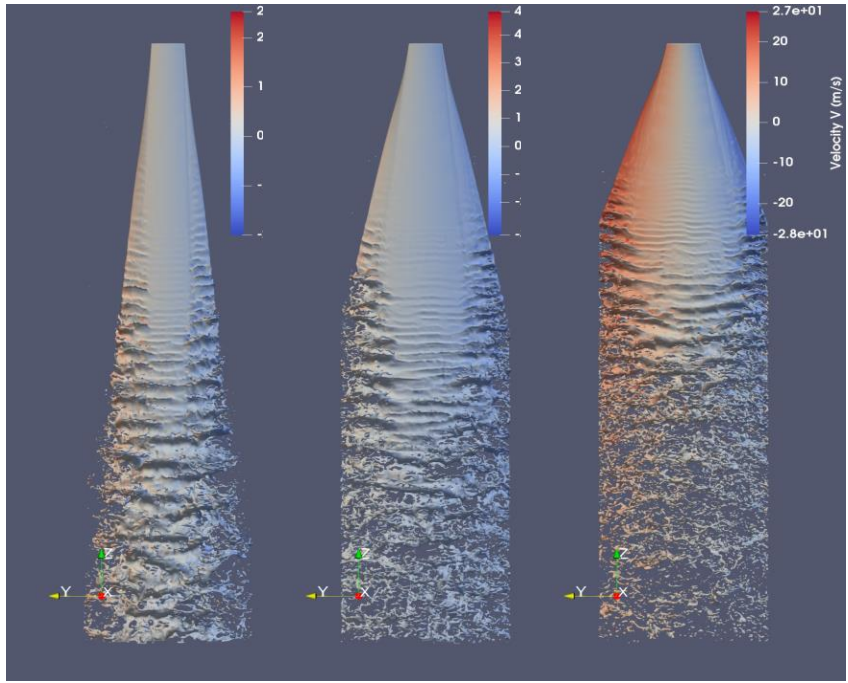

(a) 5°     (b) 10°     (c) 15°

Figure 23: *dx*, *We* = 20571. Contours of zero level set at various $\theta$.

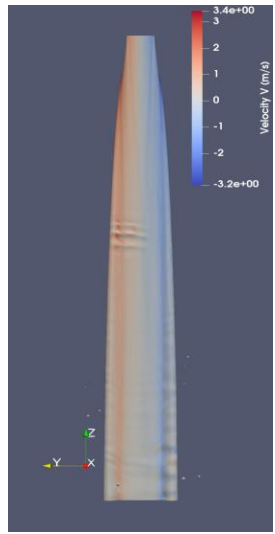
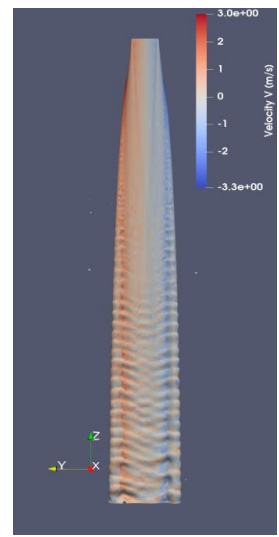

(a) $\frac{dx}{2}$     (b) *dx*

Figure 24: $\theta$ = 5°, *We* = 2285. Contours of zero level set at two resolutions.

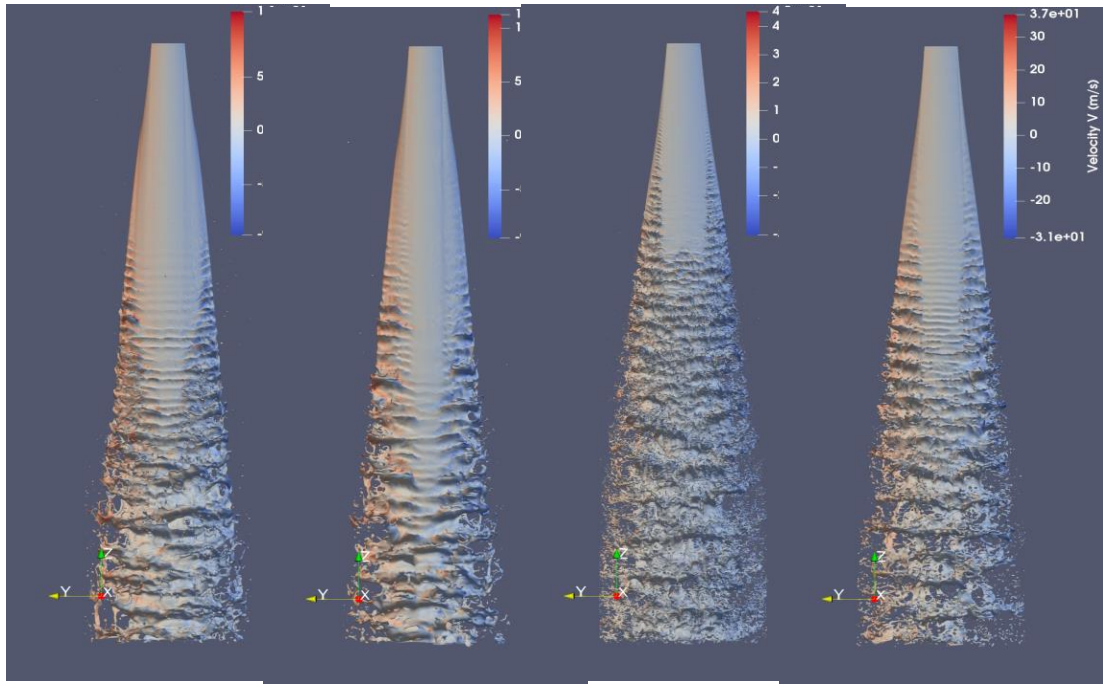

(a) $\frac{dx}{2}$, $We = 9143$    (b) *dx*, *We* = 9143    (c) $\frac{dx}{2}$, $We = 57142$    (d) *dx*, *We* = 57142

Figure 25: $\theta$ = 5°. Contours of zero level set at various Weber numbers and two resolutions.

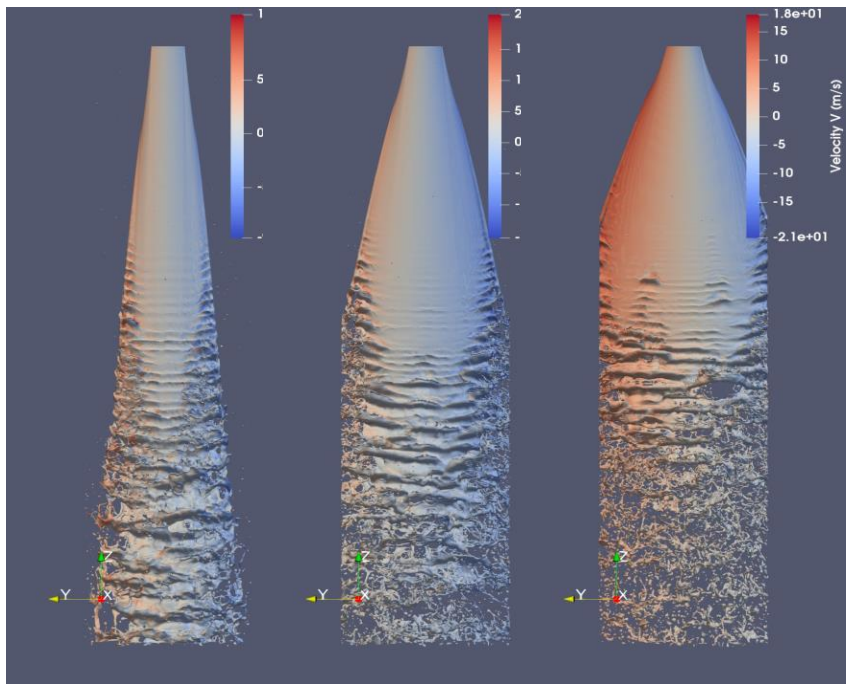

(a) $\theta$ = 5°    (b) $\theta$ = 10°    (c) $\theta$ = 15°

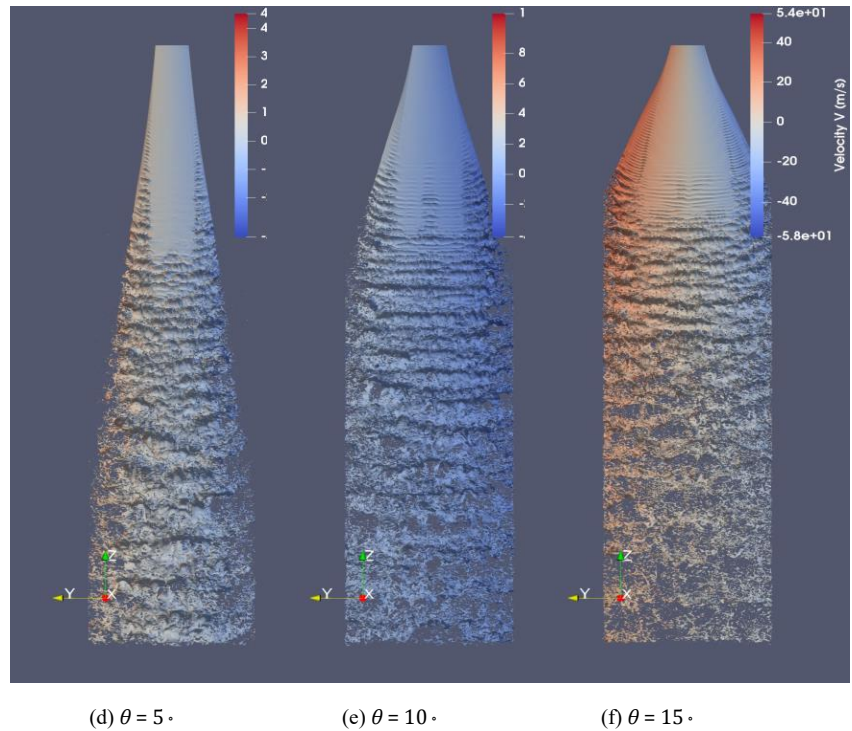

(d) $\theta = 5°$  (e) $\theta = 10°$  (f) $\theta = 15°$

Figure 26: $\frac{dx}{2}$. Contours of zero level set at various Weber numbers and $\theta$. Top row: *We* = 9143. Bottom row: *We* = 57142

# References

Agbaglah, G., Josserand, C., Zaleski, S., 2013. Longitudinal instability of a liquid rim. Physics of Fluids 25.

Ahmed, M., Amighi, A., Ashgriz, N., Tran, H., 2008. Characteristics of liquid sheets formed by splash plate nozzles. Experiments in fluids 44, 125–136.

Altieri, A., Cryer, S., Acharya, L., 2014. Mechanisms, experiment, and theory of liquid sheet breakup and drop size from agricultural nozzles. Atomization and Sprays 24, 695–721.

Altimira, M., Rivas, A., Larraona, G.S., Anton, R., Ramos, J.C., 2009. Characterization of fan spray atomizers through numerical simulation. International Journal of Heat and Fluid Flow 30, 339–355.

Asgarian, A., 2020. Physical and Mathematical Modeling of Water Atomization for Metal Powder Production. Ph.D. thesis. University of Toronto.

Asgarian, A., Heinrich, M., Schwarze, R., Bussmann, M., Chattopadhyay, K., 2020. Experiments and modeling of the breakup mechanisms of an attenuating liquid sheet. International Journal of Multiphase Flow 130, 103347.

Asuri Mukundan, A., Ménard, T., Brändle de Motta, J.C., Berlemont, A., 2022a. Detailed numerical simulations of primary atomization of airblasted liquid sheet. International Journal of Multiphase Flow 147, 103848.

Asuri Mukundan, A., Ménard, T., Brändle de Motta, J.C., Berlemont, A., 2022b. A hybrid moment of fluidlevel set framework for simulating primary atomization. Journal of Computational Physics 451, 110864.

Behzad, M., Ashgriz, N., Karney, B., 2016. Surface breakup of a nonturbulent liquid jet injected into a high pressure gaseous crossflow. International Journal of Multiphase Flow 80, 100–117.

Chai, M., Fu, Y., Zheng, S., Hong, Z., Shao, C., Luo, K., Fan, J., 2023. Detailed numerical simulation of multi-scale interface-vortex interactions of liquid jet atomization in crossflow. International Journal of Multiphase Flow 161, 104390.

Chéron, V., de Motta, J.C.B., Blaisot, J.B., Menard, T., 2022. Analysis of the effect of the 2d projection on droplet shape parameters. Atomization and Sprays 32.

Chorin, A.J., 1997. A numerical method for solving incompressible viscous flow problems. Journal of computational physics 135, 118–125.

Chéron, V., 2020. Couplage de la méthode de capture d'interface et de particules lagrangiennes pour la simulation de l'atomisation. Ph.D. thesis. Université de Rouen.

Clark, C., Dombrowski, N., 1972. Aerodynamic instability and disintegration of inviscid liquid sheets. Proceedings of the Royal Society of London. A. Mathematical and Physical Sciences 329, 467–478.

Clark, C., Dombrowski, N., 1974. An experimental study of the flow of thin liquid sheets in hot atmospheres. Journal of Fluid Mechanics 64, 167–175.

Crapper, G., Dombrowski, N., Jepson, W., Pyott, G., 1973. A note on the growth of kelvin-helmholtz waves on thin liquid sheets. Journal of Fluid Mechanics 57, 671–672.

Culick, F.E., 1960. Comments on a ruptured soap film. Journal of applied physics 31, 1128–1129.

Deberne, C., Chéron, V., Poux, A., de Motta, J.C.B., 2024a. Breakup prediction of oscillating droplets under turbulent flow. International Journal of Multiphase Flow 173, 104731.
44

Deberne, C., Renoult, M.C., Blaisot, J.B., 2024b. A mathematical model describing an unsteady leak of compressed air from an open underwater storage tank. Journal of Energy Storage 81, 110318.

Deka, H., Pierson, J.L., Soares, E.J., 2019. Retraction of a viscoplastic liquid sheet. Journal of Non-Newtonian Fluid Mechanics 272, 104172.

Dombrowski, N., Foumeny, E., 1998. On the stability of liquid sheets in hot atmospheres. Atomization and Sprays 8.

Dombrowski, N., Fraser, R.P., 1954. A photographic investigation into the disintegration of liquid sheets. Philosophical Transactions of the Royal Society of London. Series A, Mathematical and Physical Sciences 247, 101 – 130.

Dombrowski, N., Hasson, D., Ward, D., 1960. Some aspects of liquid flow through fan spray nozzles. Chemical Engineering Science 12, 35–50.

Dombrowski, N., Hooper, P.C., 1962. The effect of ambient density on drop formation in sprays. Chemical Engineering Science 17, 291–305.

Dombrowski, N., Johns, W., 1963. The aerodynamic instability and disintegration of viscous liquid sheets. Chemical Engineering Science 18, 203–214.

Duret, B., 2013. Simulation numérique directe des écoulements liquide-gaz avec évaporation : application à l'atomisation. Ph.D. thesis. Université de Rouen.

Duret, B., Canu, R., Reveillon, J., Demoulin, F., 2018. A pressure based method for vaporizing compressible two-phase flows with interface capturing approach. International Journal of Multiphase Flow 108, 42–50.

Fedkiw, R.P., Aslam, T., Merriman, B., Osher, S., 1999. A non-oscillatory eulerian approach to interfaces in multimaterial flows (the ghost fluid method). Journal of computational physics 152, 457–492.

Fraser, R.P., Eisenklam, P., Dombrowski, N., Hasson, D., 1962. Drop formation from rapidly moving liquid sheets. AIChE Journal 8, 672–680.

Fullana, J.M., Zaleski, S., 1999. Stability of a growing end rim in a liquid sheet of uniform thickness. Physics of Fluids 11, 952–954.

Gaillard, A., Sijs, R., Bonn, D., 2022. What determines the drop size in sprays of polymer solutions? Journal of Non-Newtonian Fluid Mechanics 305, 104813.

Harlow, F.H., Welch, J.E., et al., 1965. Numerical calculation of timedependent viscous incompressible flow of fluid with free surface. Physics of fluids 8, 2182.

Inamura, T., Yanaoka, H., Tomoda, T., 2004. Prediction of mean droplet size of sprays issued from wall impingement injector. AIAA journal 42, 614–621.

Jiao, D., Zhang, F., Du, Q., Niu, Z., Jiao, K., 2017. Direct numerical simulation of near nozzle diesel jet evolution with full temporal-spatial turbulence inlet profile. Fuel 207, 22–32.

Kashani, A., Parizi, H., Mertins, K., 2018. Multi-step spray modelling of a flat fan atomizer. Computers and Electronics in Agriculture 144, 58–70.

Kooij, S., Sijs, R., Denn, M.M., Villermaux, E., Bonn, D., 2018. What determines the drop size in sprays? Physical review X 8, 031019.

Li, X., Soteriou, M.C., 2018. Detailed numerical simulation of liquid jet atomization in crossflow of increasing density. International Journal of Multiphase Flow 104, 214–232.

Luret, G., 2010. Etude numérique des phénomènes de collision / coalescence en milieu dense. Ph.D. thesis. Université de Rouen.

Martinez, L.G., Duret, B., Reveillon, J., Demoulin, F., 2023. Vapor mixing in turbulent vaporizing flows. International Journal of Multiphase Flow 161, 104388.

Martinez, L.G., Duret, B., Reveillon, J., Demoulin, F.X., 2021. A new dns formalism dedicated to turbulent two-phase flows with phase change. International Journal of Multiphase Flow 143, 103762.

Ménard, T., Tanguy, S., Berlemont, A., 2007. Coupling level set/vof/ghost fluid methods: Validation and application to 3d simulation of the primary break-up of a liquid jet. International Journal of Multiphase Flow 33, 510–524.

Mukundan, A.A., Tretola, G., Ménard, T., Herrmann, M., Navarro-Martinez, S., Vogiatzaki, K., de Motta, J.C.B., Berlemont, A., 2021. Dns and les of primary atomization of turbulent liquid jet injection into a gaseous crossflow environment. Proceedings of the Combustion Institute 38, 3233–3241.

Odier, N., Balarac, G., Corre, C., Moureau, V., 2015. Numerical study of a flapping liquid sheet sheared by a high-speed stream. International Journal of Multiphase Flow 77, 196–208.

Post, S.L., Hewitt, A.J., 2018. Flat-fan spray atomization model. Transactions of the ASABE 61, 1249–1256.

Rayleigh, L., 1879. On the capillary phenomena of jets. Proceedings of the royal society of London , 71–97.

Rivas, G.S., Ramos, J.C., Altimira, J.F., 2005. Analysis of liquid-gas flow near a fan-spray nozzle outlet, in: Proceedings of the 20th ILASS-Europe Meeting.

Roa, I., Renoult, M.C., Dumouchel, C., Brändle de Motta, J.C., 2023. Droplet oscillations in a turbulent flow. Frontiers in Physics 11, 1173521.

Rudman, M., 1998. A volume-tracking method for incompressible multifluid flows with large density variations. International Journal for numerical methods in fluids 28, 357–378.

Sanadi, D.S., 2022. Analysis of liquid sheet thickness and perforation kinematics using conventional and time-gated optical diagnostics. Ph.D. thesis. Normandie Université.

Sarchami, A., Ashgriz, N., Tran, H., 2009. Initial perturbation amplitude of liquid sheets produced by jet-impingement nozzles. AIAA journal 47, 2775–2779.

Senecal, P., Schmidt, D., Nouar, I., Rutland, C., Reitz, R., Corradini, M., 1999. Modeling high-speed viscous liquid sheet atomization. International Journal of Multiphase Flow 25, 1073–1097.

Squire, H., 1953. Investigation of the instability of a moving liquid film. British Journal of applied physics 4, 167.

Sussman, M., Smith, K.M., Hussaini, M.Y., Ohta, M., Zhi-Wei, R., 2007. A sharp interface method for incompressible two-phase flows. Journal of computational physics 221, 469–505.

Tanguy, S., Ménard, T., Berlemont, A., 2007. A level set method for vaporizing two-phase flows. Journal of Computational Physics 221, 837–853.

Taylor, G.I., 1959. The dynamics of thin sheets of fluid. iii. disintegration of fluid sheets. Proceedings of the Royal Society of London. Series A. Mathematical and Physical Sciences 253, 313–321.

Vaudor, G., Ménard, T., Aniszewski, W., Doring, M., Berlemont, A., 2017. A consistent mass and momentum flux computation method for two phase flows. application to atomization process. Computers & Fluids 152, 204– 216.

Villermaux, E., Clanet, C., 2002. Life of a flapping liquid sheet. Journal of fluid mechanics 462, 341–363.

Weber, C., 1931. Zum zerfall eines flüssigkeitsstrahles. ZAMM - Journal of Applied Mathematics and Mechanics / Zeitschrift für Angewandte Mathematik und Mechanik 11, 136–154.